\PassOptionsToPackage{unicode}{hyperref}
\PassOptionsToPackage{hyphens}{url}
\PassOptionsToPackage{dvipsnames,svgnames,x11names}{xcolor}
\documentclass[
  12pt]{article}

\usepackage{amsthm,amsmath,amsfonts,amssymb, mathtools, physics}
\usepackage{xypic}
\usepackage{xcolor}
\usepackage{natbib}
\usepackage{algorithm}
\usepackage{algpseudocode}
\usepackage{comment}
\usepackage{lipsum}
\usepackage{subcaption}
\usepackage{multirow}
\usepackage{chngcntr}
\usepackage{enumitem}
\usepackage{kotex}
\usepackage{filecontents}

\DeclareMathOperator*{\argmax}{arg\,max}

\newtheorem{thm}{Theorem}
\newtheorem{lem}[thm]{Lemma} 
\newtheorem{prop}[thm]{Proposition}
\newtheorem{remark}{Remark}

\newtheorem{assumption}{Assumption}

\newcommand{\SigX}{\Sigma_X}
\newcommand{\SigY}{\Sigma_Y}

\newcommand{\hrS}{\hat{r}_S}

\newcommand{\QS}{Q_S}
\newcommand{\QD}{Q_D}
\newcommand{\US}{U_S}
\newcommand{\UD}{U_D}

\newcommand{\hQ}{\hat{Q}}
\newcommand{\hU}{\hat{U}}
\newcommand{\hP}{\hat{P}}
\newcommand{\hq}{\hat{q}}
\newcommand{\hu}{\hat{u}}

\newcommand{\hlamb}{\hat{\lambda}}
\newcommand{\hgamm}{\hat{\gamma}}
\newcommand{\tlamb}{\tilde{\lambda}}
\newcommand{\tgamm}{\tilde{\gamma}}

\newcommand{\hPsiX}{\hat{\Psi}_X}
\newcommand{\hPsiY}{\hat{\Psi}_Y}

\newcommand{\LambS}{\Lambda_S}
\newcommand{\GammS}{\Gamma_S}
\newcommand{\LambD}{\Lambda_D}
\newcommand{\GammD}{\Gamma_D}

\newcommand{\Real}{\mathbb{R}}

\newcommand{\aslim}{\overset{\text{a.s.}}{\rightarrow}}

\newcommand{\iid}{\overset{\text{i.i.d.}}{\sim}}
\newcommand{\inner}[2]{\langle #1, #2 \rangle}
\newcommand{\calO}{\mathcal{O}}

\newcommand{\calZ}{\mathcal{Z}}
\newcommand{\calW}{\mathcal{W}}

\newcommand{\tspan}{\text{span}}

\usepackage{xr}

\makeatletter
\newcommand*{\addFileDependency}[1]{
  \typeout{(#1)}
  \@addtofilelist{#1}
  \IfFileExists{#1}{}{\typeout{No file #1.}}
}
\makeatother

\ifdefined\proof
    \renewenvironment{proof}{\par\noindent{\bf Proof\ }}{\hfill\BlackBox\\[2mm]}
\else
    \newenvironment{proof}{\par\noindent{\bf Proof\ }}{\hfill\BlackBox\\[2mm]}
\fi

\usepackage{iftex}
\ifPDFTeX
  \usepackage[T1]{fontenc}
  \usepackage[utf8]{inputenc}
  \usepackage{textcomp} 
\else 
  \usepackage{unicode-math}
  \defaultfontfeatures{Scale=MatchLowercase}
  \defaultfontfeatures[\rmfamily]{Ligatures=TeX,Scale=1}
\fi
\usepackage{lmodern}
\ifPDFTeX\else  
\fi
\IfFileExists{upquote.sty}{\usepackage{upquote}}{}
\IfFileExists{microtype.sty}{
  \usepackage[]{microtype}
  \UseMicrotypeSet[protrusion]{basicmath} 
}{}
\makeatletter
\@ifundefined{KOMAClassName}{
  \IfFileExists{parskip.sty}{%
    \usepackage{parskip}
  }{
    \setlength{\parindent}{0pt}
    \setlength{\parskip}{6pt plus 2pt minus 1pt}}
}{
  \KOMAoptions{parskip=half}}
\makeatother
\usepackage{xcolor}
\makeatletter
\ifx\paragraph\undefined\else
  \let\oldparagraph\paragraph
  \renewcommand{\paragraph}{
    \@ifstar
      \xxxParagraphStar
      \xxxParagraphNoStar
  }
  \newcommand{\xxxParagraphStar}[1]{\oldparagraph*{#1}\mbox{}}
  \newcommand{\xxxParagraphNoStar}[1]{\oldparagraph{#1}\mbox{}}
\fi
\ifx\subparagraph\undefined\else
  \let\oldsubparagraph\subparagraph
  \renewcommand{\subparagraph}{
    \@ifstar
      \xxxSubParagraphStar
      \xxxSubParagraphNoStar
  }
  \newcommand{\xxxSubParagraphStar}[1]{\oldsubparagraph*{#1}\mbox{}}
  \newcommand{\xxxSubParagraphNoStar}[1]{\oldsubparagraph{#1}\mbox{}}
\fi
\makeatother

\usepackage{longtable,booktabs,array}
\usepackage{calc} 
\usepackage{etoolbox}
\makeatletter
\patchcmd\longtable{\par}{\if@noskipsec\mbox{}\fi\par}{}{}
\makeatother
\IfFileExists{footnotehyper.sty}{\usepackage{footnotehyper}}{\usepackage{footnote}}
\makesavenoteenv{longtable}
\usepackage{graphicx}
\makeatletter
\def\maxwidth{\ifdim\Gin@nat@width>\linewidth\linewidth\else\Gin@nat@width\fi}
\def\maxheight{\ifdim\Gin@nat@height>\textheight\textheight\else\Gin@nat@height\fi}
\makeatother
\setkeys{Gin}{width=\maxwidth,height=\maxheight,keepaspectratio}
\makeatletter
\def\fps@figure{htbp}
\makeatother

\makeatletter
\@ifpackageloaded{caption}{}{\usepackage{caption}}
\AtBeginDocument{%
\ifdefined\contentsname
  \renewcommand*\contentsname{Table of contents}
\else
  \newcommand\contentsname{Table of contents}
\fi
\ifdefined\listfigurename
  \renewcommand*\listfigurename{List of Figures}
\else
  \newcommand\listfigurename{List of Figures}
\fi
\ifdefined\listtablename
  \renewcommand*\listtablename{List of Tables}
\else
  \newcommand\listtablename{List of Tables}
\fi
\ifdefined\figurename
  \renewcommand*\figurename{Figure}
\else
  \newcommand\figurename{Figure}
\fi
\ifdefined\tablename
  \renewcommand*\tablename{Table}
\else
  \newcommand\tablename{Table}
\fi
}
\@ifpackageloaded{float}{}{\usepackage{float}}
\floatstyle{ruled}
\@ifundefined{c@chapter}{\newfloat{codelisting}{h}{lop}}{\newfloat{codelisting}{h}{lop}[chapter]}
\floatname{codelisting}{Listing}

\makeatother
\makeatletter
\@ifpackageloaded{caption}{}{\usepackage{caption}}
\@ifpackageloaded{subcaption}{}{\usepackage{subcaption}}
\makeatother

\ifLuaTeX
  \usepackage{selnolig}  
\fi
\usepackage[]{natbib}
\usepackage{bookmark}

\IfFileExists{xurl.sty}{\usepackage{xurl}}{} 
\hypersetup{
  pdftitle={Title},
  pdfauthor={Author 1; Author 2},
  pdfkeywords={3 to 6 keywords, that do not appear in the title},
  colorlinks=true,
  linkcolor={blue},
  filecolor={Maroon},
  citecolor={Blue},
  urlcolor={Blue},
  pdfcreator={LaTeX via pandoc}}

\newcommand{\anon}{1}

\begin{document}

\def\spacingset#1{\renewcommand{\baselinestretch}%
{#1}\small\normalsize} \spacingset{1}


\if1\anon
{
  \title{\bf Joint estimation of high-dimensional spiked covariance matrices via a partially shared subspace}
\author{Changwon Yoon \\ Department of Industrial and Systems Engineering, KAIST \\ Minwoo Kim \\ Institute for Data Innovation in Science, Seoul National University \\ Sungkyu Jung \\ Department of Statistics, Seoul National University \\ and \\ Jeongyoun Ahn\thanks{Corresponding author (jyahn@kaist.ac.kr)} \\ Department of Industrial and Systems Engineering, KAIST}
\date{\vspace{-5ex}}
  \maketitle
  
} \fi

\if0\anon
{
  \bigskip
  \bigskip
  \bigskip
  \begin{center}
    {\LARGE\bf Joint estimation of high-dimensional spiked covariance matrices via a partially shared subspace}
\end{center}
  \medskip
} \fi

\bigskip
\begin{abstract}

Statistical analysis of high-dimensional data is often hampered by limited sample sizes, yet auxiliary datasets from related sources are often readily available. When two such datasets share part of their covariance structure, but not all of it, exploiting the shared part can substantially improve estimation.  We propose a spiked covariance model that explicitly captures this partial sharing: two datasets share a subspace of unknown rank and arbitrary position in the spectrum, while each retains its own distinct spiked directions. The model treats the two datasets symmetrically and strictly generalizes existing models for shared covariance structure. We develop a complete estimation procedure that includes joint estimation of the shared subspace and its rank, a closed-form pooling weight for combining the two datasets, and asymptotic guarantees derived from random matrix theory in the proportional-growth regime. The framework also resolves a gap in contrastive dimension reduction by providing a principled estimator for high-dimensional settings. We illustrate the methodology on portfolio construction during the early COVID-19 pandemic and on contrastive analysis of brain tumor gene expression.
\end{abstract}

\noindent%
\vfill

\newpage
\spacingset{1.8} 

\section{Introduction}
\label{sec:intro}

Covariance matrices from related datasets frequently share a portion of their spiked structure. This shared component forms a common subspace that does not necessarily manifest in the leading spikes of either dataset, nor must it be expressed in the same eigenvector basis. Empirical evidence for such partial overlap is well-documented. For instance, \citet{lenz2016principal} demonstrated that the first three principal components of two independently collected gene expression datasets span nearly identical subspaces. Similar structural sharing arises in finance, where stock returns during crises and tranquil periods share broad market factors but diverge in sector-specific behaviors, and in cancer genomics, where tumor expressions and matched normal tissues share organ-level profiles but deviate along disease-specific pathways. In these settings, one dataset typically serves as the primary object of analysis (the \emph{target}) while the other provides auxiliary information (the \emph{background}). The central challenge lies in leveraging this background dataset to improve the estimation of the target's covariance matrix, effectively exploiting their shared architecture without introducing bias due to their discrepancies.

This problem has been extensively studied under the framework of \emph{common principal component analysis} \citep{flury1984common, schott1991some, wang2021semiparametric}, where covariance matrices across multiple groups are modeled as sharing a specific eigenstructure. However, classical approaches typically assume that the entire set of eigenvectors is shared, that the shared directions strictly constitute the leading spikes, or that the feature dimension remains small relative to the sample size — assumptions that are frequently violated simultaneously in modern high-dimensional applications. In this paper, we develop a framework that relaxes all three limitations. Specifically, we introduce a spiked covariance model in which the target and background covariance matrices share a \emph{subspace} of unknown rank and \emph{arbitrary} spectral position, while allowing each dataset to retain its own \emph{distinct} spiked directions. 

We adopt the spiked covariance framework \citep{johnstone2001distribution} under the high-dimensional, proportional-growth regime where $p, n_X, n_Y \to \infty$ such that $p/n_X \to c_X \in (0,\infty)$ and $p/n_Y \to c_Y \in (0, \infty)$. Here, $p$ denotes the dimension size while $n_X$ and $n_Y$ are the sample sizes of target and background datasets respectively. Our proposed \emph{partially shared subspace} (PSS) model decomposes the spiked eigenspace of each covariance matrix into two orthogonal structural modules: a shared component that spans the common subspace across both datasets, and a distinct component unique to each. The primary contributions of this work are organized as follows:

First, we provide a principled procedure for identifying the shared subspace itself. In the high-dimensional proportional-growth regime, the sample spiked eigenvectors are known to be inconsistent for their population counterparts; instead, the inner products between sample and population eigenvectors converge to deterministic limits strictly bounded between zero and one \citep{paul2007asymptotics}. By analyzing the principal angles between the sample spiked subspaces of the two datasets, we fully characterize their joint asymptotic behavior. We then leverage these random matrix theory results to develop an iterative algorithm that jointly estimates both the dimension of the shared subspace and the exact indices of the spiked eigenvectors that span it.

Second, we establish an optimal framework for merging the two datasets to construct the target covariance estimator itself. We introduce a plug-in estimator based on orthogonal projections onto the shared subspace, structured as a weighted combination of the individual spiked projection operators. In this setting, we derive the closed-form pooling weight that minimizes the asymptotic Frobenius loss, revealing that the optimal balance is governed jointly by the relative sample sizes and the relative population spike magnitudes. We also quantify, in closed form, the asymptotic efficiency gain of the pooled shared subspace estimator from incorporating background data, relative to traditional target-only estimation. This improvement has broad practical relevance, as enhancements to the target covariance matrix are directly related to the performance of downstream applications such as classification, factor modeling, and dimension reduction.

Third, we address a critical theoretical gap in contrastive dimension reduction (CDR) \citep{abid2018exploring, severson2019unsupervised, zhang2025contrastive}. Inherently, our model's shared-distinct decomposition isolates a target-specific distinct subspace that captures variations absent in the background. This architecture is ideal for applications requiring structural isolation, such as separating task-induced brain activation from baseline neural noise, or distinguishing true treatment effects from control-arm patient heterogeneity. While existing CDR methods are heavily reliant on low-dimensional heuristics, such as sample covariance matrix stabilization, they lack rigorous performance guarantees when features scale proportionally with sample size. The PSS framework resolves this limitation, delivering what is, to our knowledge, the first contrastive dimensionality framework backed by asymptotic guarantees in the high-dimensional regime.

The remainder of the paper is structured as follows. Following a review of related work and notation, Section~\ref{sec:model} establishes the PSS model in the high-dimensional spiked covariance framework alongside its asymptotic foundations. Section~\ref{sec:estimation} presents our estimation methodology—including the joint rank-and-index identification algorithm, the plug-in covariance estimator, and the closed-form optimal pooling weight—with theoretical guarantees under proportional growth. Numerical simulations under model configurations and misspecifications are detailed in Section~\ref{sec:simulation}. Section~\ref{sec:real data analysis} applies the methodology to early-pandemic portfolio construction and contrastive genomics analysis of low-grade gliomas, and Section~\ref{sec:discussion} concludes the paper.

\subsubsection*{Related work}\label{sec:related works}

\textit{Common principal components.} The literature on common eigenstructures across multiple covariance matrices has progressively relaxed restrictive modeling assumptions. \citet{flury1984common} introduced classical common principal component analysis (CPCA) under the most rigid form of sharing, requiring all $p$ eigenvectors to be identical across $K$ groups while allowing only the eigenvalues to be group-specific. \citet{flury1987two}  relaxed this constraint to a partial framework where only a subset of $q < p$ eigenvectors is shared. Later \citet{schott1988common, schott1991some} shifted focus from strict eigenvector equality to subspace equality, requiring only that the leading $m$-dimensional subspaces coincide across groups. More recently, \citet{wang2021semiparametric} and \citet{shi2024personalized} accommodated the sharing of non-leading eigenvectors, while \citet{franks2019shared} extended partial common subspaces to high-dimensional regimes using a Bayesian formulation. However, the latter approach assumes no group-specific spiked structure beyond the shared component. Our model addresses these collective limitations by adopting a fully general form: the shared subspace may have an unknown rank and can consist of eigenvectors positioned at arbitrary locations within each dataset's spike spectrum.

\noindent \textit{Transfer learning for high-dimensional matrices.} Our methodology intersects with transfer learning, where an auxiliary background dataset is leveraged to enhance the statistical efficiency of target parameter estimation. Most existing transfer learning techniques for large matrices rely heavily on proximity conditions, requiring the target and source parameters to be close under a specific matrix norm or imposing strict sparsity bounds on their contrast matrix. This includes frameworks developed for Gaussian graphical models \citep{li2023transfer, ren2024transfer}, principal component analysis \citep{li2024knowledge, hendy2024tl}, and direct covariance estimation \citep{cai2024transfer}. Rather than requiring bounded differences in parameters, we pose a purely geometric condition: the two covariance matrices share a low-dimensional subspace but are otherwise unconstrained. This structural formulation safeguards the model against misspecification; if the background dataset shares no common structure with the target, the estimated shared dimension collapses to zero, effectively eliminating negative transfer.

\noindent \textit{Contrastive dimension reduction.} Our framework also intersects with contrastive dimension reduction (CDR), a recent line of work that addresses the same core objective as the PSS distinct subspace: isolating directions of variation present in a target dataset but absent from a related background dataset. The pioneering method, contrastive PCA \citep{abid2018exploring}, maximizes $v^\top (S_X - \alpha S_Y) v$ over unit vectors $v$, where $S_X$ and $S_Y$ are sample covariance matrices of target and background data, respectively, while $\alpha$ acts as a user-specified contrastive parameter. Subsequent literature has refined this paradigm in two primary directions: model-based objectives \citep{severson2019unsupervised, li2024probabilistic} and generalized objectives that circumvent the contrastive tuning parameter entirely \citep{de2025identifying}. For a comprehensive overview of this landscape, see \citet{hawke2025contrastive}. However, these approaches operate almost exclusively under low-dimensional paradigms where sample covariance matrices are stable. The distinct subspace from our PSS procedure yields a principled CDR estimator, establishing theoretical consistency in high-dimensional regimes where existing CDR methods offer no statistical guarantees.

\subsubsection*{Notations}

We establish the notation used throughout the paper. Let $\aslim$ denote almost sure convergence. We write $\Real^{n \times m}$ for the set of $n \times m$ real matrices and $\calO^{n \times m}$ for the set of $n \times m$ semi-orthogonal matrices ($A^\top A = I_m$ with $n \ge m$). The zero matrix and zero vector are written as $0_{n \times m}$ and $0_d$, respectively. For $A \in \Real^{n \times m}$ and $B \in \Real^{n \times l}$, $A \perp B$ indicates that $A^\top B = 0_{m \times l}$. Let $\mathrm{diag}(a_1,\ldots,a_d)$ be the diagonal matrix with entries $a_1,\ldots,a_d$, and let $I_d$ be the $d \times d$ identity matrix. For $A \in \Real^{n \times m}$, $\sigma_i(A)$ is its $i$-th largest singular value and $\sigma(A) = \{\sigma_i(A)\}_{i=1}^{\min(n,m)}$ is its singular value multiset; its Frobenius norm is $\|A\|_F$. For a symmetric positive semi-definite matrix $\Sigma \in \Real^{n \times n}$, $\ell_i(\Sigma)$ and $\ell(\Sigma) = \{\ell_i(\Sigma)\}_{i=1}^{n}$ denote its $i$-th largest eigenvalue and its eigenvalue multiset, respectively. For vectors $u, v \in \Real^d$, $\langle u, v \rangle$ represents the standard inner product, and $\|u\|$ represents the $L_2$ norm.

\section{Spiked covariance model with partially shared subspace}\label{sec:model}

This section presents our primary modeling framework. Section~\ref{sec:setup} defines the two-dataset configuration and the spiked covariance model characterizing each dataset, while Section~\ref{sec:pcsmodel} introduces the partially shared subspace structure that couples them.

\subsection{Setup}\label{sec:setup}

We consider two independent datasets comprising a common set of $p$ variables, denoted by $\{X_i\}_{i=1}^{n_X}$ and $\{Y_i\}_{i=1}^{n_Y}$, where $X_i \iid \mathcal{N}(0_p,\SigX)$ and $Y_i \iid \mathcal{N}(0_p,\SigY)$. In typical applications, one dataset is of primary interest while the other serves as an auxiliary source of information; we refer to these as the \emph{target} and \emph{background} datasets, respectively, and maintain this terminology throughout. We emphasize, however, that our proposed model and estimation procedure treat the two datasets symmetrically: these labels merely reflect the analyst's focus rather than any underlying structural asymmetry. The respective sample sizes $n_X$ and $n_Y$ are completely unconstrained and frequently unbalanced. In clinical or observational settings, such as a disease-specific target cohort paired with a large general-population background control, $n_Y$ often heavily dominates $n_X$. Because our primary objective is to characterize and exploit the shared covariance structure, we assume zero-mean vectors without loss of generality.

We assume that $\SigX$ and $\SigY$ follow a spiked covariance model,
\[
\SigX = Q \Lambda Q^\top + \tau^2 I_p
\qquad \text{and} \qquad
\SigY = U \Gamma U^\top + \eta^2 I_p,
\]
where $Q = [q_1,\ldots,q_{r_X}] \in \calO^{p \times r_X}$ and $U = [u_1,\ldots,u_{r_Y}] \in \calO^{p \times r_Y}$ collect the spiked eigenvectors, and
$\Lambda = \mathrm{diag}(\lambda_1 - \tau^2,\ldots,\lambda_{r_X} - \tau^2), \;
\Gamma = \mathrm{diag}(\gamma_1 - \eta^2,\ldots,\gamma_{r_Y} - \eta^2),$
with $\lambda_1 > \cdots > \lambda_{r_X} > \tau^2$ and $\gamma_1 > \cdots > \gamma_{r_Y} > \eta^2$. Under this parametrization, the eigenvalues of the two covariance matrices are
\[
\ell(\SigX) = \{\lambda_1,\ldots,\lambda_{r_X},
              \underbrace{\tau^2,\ldots,\tau^2}_{p - r_X}\}
\quad \text{and} \quad
\ell(\SigY) = \{\gamma_1,\ldots,\gamma_{r_Y},
              \underbrace{\eta^2,\ldots,\eta^2}_{p - r_Y}\},
\]
so that $\lambda_1,\ldots,\lambda_{r_X}$ and $\gamma_1,\ldots,\gamma_{r_Y}$ are the spikes of $\SigX$ and $\SigY$, while $\tau^2$ and $\eta^2$ are their respective bulk (noise) levels. The spikes are assumed to be distinct for simplicity.

We denote the sample covariance matrices by $S_{X} = n_X^{-1}\sum_{i=1}^{n_X} X_i X_i^\top$ and $S_Y = n_Y^{-1}\sum_{i=1}^{n_Y} Y_i Y_i^\top$. The matrices $\hQ = [\hq_1,\ldots,\hq_{r_X}]$ and $\hU = [\hu_1,\ldots,\hu_{r_Y}]$ collect the sample eigenvectors corresponding to their population counterparts $Q$ and $U$ , while the sample eigenvalues corresponding to $\lambda_i$ and $\gamma_j$ are denoted by $\hlamb_i$ and $\hgamm_j$ for $i = 1,\ldots,r_X$ and $j = 1,\ldots,r_Y$. In the high-dimensional regime, the spike counts $r_X, r_Y$ and the baseline noise levels $\tau^2, \eta^2$ can be consistently estimated using well-established algorithms, such as the bulk eigenvalue matching analysis proposed by \citet{ke2023estimation}, which we adopt in all our empirical implementations. Consequently, we treat $r_X$ and $r_Y$ as known parameters throughout the development of our methodology, and, without loss of generality, rescale the data to set the bulk noise levels to $\tau^2 = \eta^2 = 1$.

\subsection{The partially shared subspace model}\label{sec:pcsmodel}

The auxiliary background data can improve the estimation of the target covariance matrix only insofar as the two populations share underlying structure. We assume this sharing manifests as the signal directions of $\SigX$ and $\SigY$ spanning a common subspace, while the remaining directions are specific to each dataset.

Formally, we assume that $\SigX$ and $\SigY$ admit the decompositions
\begin{equation}
    \label{eq:PSS}
    \begin{aligned}        
\SigX &= \QS \LambS \QS^\top+ \QD \LambD \QD^\top+ I_p,\\
\SigY &= \US \GammS \US^\top+ \UD \GammD \UD^\top+ I_p,
    \end{aligned}
\end{equation}
partitioning the spiked eigenvectors into shared and distinct components. We call this the partially shared subspace (PSS) model. The shared bases $\QS, \US \in \calO^{p \times r_S}$ span a common $r_S$-dimensional subspace such that $\US = \QS R$ for an $r_S \times r_S$ rotation matrix $R$. Conversely, the distinct bases $\QD \in \calO^{p \times d_X}$ and $\UD \in \calO^{p \times d_Y}$ span the target- and background-specific subspaces. The shared subspace rank $r_S$ serves as the central parameter of interest, yielding $r_X = r_S + d_X$ and $r_Y = r_S + d_Y$. Since $[\QS, \QD]$ and $[\US, \UD]$ form orthonormal bases, intra-dataset orthogonality requires $\QS \perp \QD$ and $\US \perp \UD$. The rotational linkage $\US = \QS R$ further enforces the cross-orthogonality relations $\QS \perp \UD$ and $\US \perp \QD$. The distinct subspaces $\mathrm{span}(\QD)$ and $\mathrm{span}(\UD)$ remain otherwise unconstrained.

Crucially, we do not require the shared spikes to be the \emph{leading} ones; they may be arbitrarily interleaved across the spectrum. Let $\Psi_X \subseteq \{1,\ldots,r_X\}$ and $\Psi_Y \subseteq \{1,\ldots,r_Y\}$ denote the index sets of the spiked eigenvectors of $\SigX$ and $\SigY$ that span the shared subspace, such that $\QS$ and $\US$ contain the columns of $Q$ and $U$ indexed by $\Psi_X$ and $\Psi_Y$ (where $|\Psi_X| = |\Psi_Y| = r_S$). Explicitly, we write these columns as
\[
\begin{aligned}
\QS &= [q_{1,S},\ldots,q_{r_S,S}], &
\QD &= [q_{1,D},\ldots,q_{d_X,D}], \\
\US &= [u_{1,S},\ldots,u_{r_S,S}], &
\UD &= [u_{1,D},\ldots,u_{d_Y,D}],
\end{aligned}
\]
where $\lambda_{i,S}, \lambda_{i,D}$ denote the eigenvalues of $\SigX$ corresponding to $q_{i,S}$ and $q_{i,D}$, and $\gamma_{j,S}, \gamma_{j,D}$ denote the eigenvalues of $\SigY$ corresponding to $u_{j,S}$ and $u_{j,D}$. Permitting the shared spikes to lie anywhere in the spectrum provides a significantly more relaxed positional assumption than standard common principal component frameworks, which generally assume shared directions constitute the leading spikes.

\begin{remark} \label{rmk:franks}
The two closely related prior models are those of \citet{franks2019shared} and \citet{shi2024personalized}. The former considers covariance matrices of $K$ groups expressed as $
\Sigma_k = V \Omega_k V^\top+ I_p, \; k = 1,\ldots,K,
$
where $V\in \calO^{p\times d}$, $\Omega_k \in \mathbb R^{d\times d}$ is a positive definite matrix. The latter assumes
$
\Sigma_k = U\Phi U^\top + V_{k}\Theta_k V_{k}^\top + \sigma^2 I, \; k = 1, \ldots,K,
$
where $U \in \calO^{p\times r_1}$ is shared across groups, $V_{k} \in \calO^{p\times r_{2,k}}$ are group-specific, and $\Phi, \Theta_k$ are diagonal. When $K=2$, both models reduce to special cases of the PSS model in \eqref{eq:PSS}.
\end{remark}

\section{Estimating the PSS model}\label{sec:estimation}

Having defined the PSS model, we now turn to estimation. We first outline the asymptotic results for the sample eigenstructure that underpin our approach (Section~\ref{sec:asymptotics}). We then present a procedure for estimating the shared subspace rank and index sets (Section~\ref{sec:rank}), and conclude by constructing plug-in estimators for the target covariance matrix (Section~\ref{sec:covariance}). Throughout, the estimated eigenvectors and eigenvalues are denoted by $\hQ_S, \hQ_D, \hU_S, \hU_D$ (with columns $\hq_{i,S}$, etc.), and $\hlamb_{i,S}, \hgamm_{j,S}$, respectively. 

\subsection{Asymptotics of sample eigenstructure}\label{sec:asymptotics}

Our estimation procedures rest on the high-dimensional asymptotic behavior of sample eigenstructures. We first review the single-sample spiked covariance model \citep{johnstone2001distribution} and then extend it to our two-sample setting. Let $\{X_i\}_{i=1}^n$ be $p$-dimensional random vectors sampled i.i.d. from $\mathcal{N}(0_p,\Sigma)$. We assume $\Sigma$ follows a spiked covariance model with $r$ distinct spikes and unit bulk eigenvalues:
\[
\ell_1 > \cdots > \ell_r > \ell_{r+1} = \cdots = \ell_p = 1.
\]
Let $(v_i,\ell_i)$ denote the $i$-th eigenvector--eigenvalue pair of $\Sigma$. For the sample covariance matrix $S_n = (1/n)\sum_{i=1}^{n} X_i X_i^\top $, we denote the corresponding pair by $(\hat v_{i,n}, \hat\ell_{i,n})$.

We consider the high-dimensional regime where $p/n \to c$ as $n\rightarrow \infty$ for some $0 < c < \infty$, with both the number of spikes $r$ and their magnitudes remaining fixed. Here, a population spike must exceed the critical threshold $1 + \sqrt{c}$ for its sample counterpart to retain information about the underlying signal; below the threshold, the sample spike is asymptotically indistinguishable from the noise bulk. This phenomenon is the Baik--Ben~Arous--P\'ech\'e (BBP) phase transition \citep{baik2006eigenvalues, paul2007asymptotics, johnstone2018pca}, detailed below:

\begin{prop}\label{prop:classic asymptotics}
    Suppose $p/n \to c \in (0,\infty)$. Then, for $1 \le i, j \le r$,
    \[
    |\inner{v_i}{\hat{v}_{j,n}}| \aslim
    \begin{cases}
        \phi_c(\ell_i)\,\delta_{i,j}, & \ell_i > 1 + \sqrt{c}, \\[2pt]
        0, & \ell_i \le 1 + \sqrt{c},
    \end{cases}
    \quad
    \hat{\ell}_{i,n} \aslim
    \begin{cases}
        \ell_i + c\,\ell_i/(\ell_i - 1), & \ell_i > 1 + \sqrt{c}, \\[2pt]
        (1 + \sqrt{c})^2, & \ell_i \le 1 + \sqrt{c},
    \end{cases}
    \]
    as $n \to \infty$, where $\delta_{i,j}$ is the Kronecker delta and
    $\phi_c(\ell) = \sqrt{(1 - c/(\ell-1)^2)\,/\,(1 + c/(\ell-1))}$
    for $\ell > 1 + \sqrt{c}$.
\end{prop}


We now extend Proposition~\ref{prop:classic asymptotics} to the two-sample setting central to this paper, where the key quantity of interest is the inner product between sample eigenvectors derived from two independent datasets. Let the datasets $\{X_i\}_{i=1}^{n_X}$ and $\{Y_i\}_{i=1}^{n_Y}$, their population covariance matrices $\SigX$ and $\SigY$, and the sample eigenvectors $\hat q_i$ and $\hat u_j$ be defined as in Section~\ref{sec:setup}. The following lemma characterizes the limiting behavior of the inner product $\langle \hat q_i, \hat u_j\rangle $.

\begin{lem}\label{lem:sample inner product}
    Suppose $p/n_X \to c_X \in (0,\infty)$ and $p/n_Y \to c_Y \in (0,\infty)$,
    and that $\lambda_{r_X} > 1 + \sqrt{c_X}$ and
    $\gamma_{r_Y} > 1 + \sqrt{c_Y}$. Then, for $1 \le i \le r_X$ and
    $1 \le j \le r_Y$,
    \[
    |\inner{\hat q_i}{\hat u_j}| \aslim
    \phi_{c_X}(\lambda_i)\,\phi_{c_Y}(\gamma_j)\,|\inner{q_i}{u_j}|,
    \qquad \text{as } n_X, n_Y \to \infty.
    \]
\end{lem}

Lemma~\ref{lem:sample inner product} establishes the two-sample counterpart of Proposition~\ref{prop:classic asymptotics}: the inner product between two sample eigenvectors converges to the inner product of the corresponding population eigenvectors, scaled by the factors $\phi_{c_X}$ and $\phi_{c_Y}$. This result yields two immediate consequences that we utilize repeatedly. First, when the population eigenvectors $q_i$ and $u_j$ are orthogonal, the limiting sample inner product is zero; thus, population orthogonality is preserved asymptotically. Second, because $\phi_c(\cdot) < 1$, the sample inner product is strictly attenuated relative to its population value, meaning its magnitude never exceeds the population inner product asymptotically. These properties underpin the methodological developments in the remainder of this section.

As seen from Lemma~\ref{lem:sample inner product}, our theoretical results are developed under the following two regularizing assumptions:

\begin{assumption}\label{ass:high-dim regime}
The dimension $p$ grows proportionally with both sample sizes, satisfying $p/n_X \to c_X \in (0,\infty)$ and $p/n_Y \to c_Y \in (0,\infty)$ as $n_X, n_Y \to \infty$.
\end{assumption}

\begin{assumption}\label{ass:above transition}
All population spikes strictly exceed their respective phase transition thresholds: $\lambda_{r_X} > 1 + \sqrt{c_X}$ and $\gamma_{r_Y} > 1 + \sqrt{c_Y}$.
\end{assumption}

Assumption~\ref{ass:high-dim regime} specifies the high-dimensional regime in which $p$ and the sample sizes diverge together. Assumption~\ref{ass:above transition} places all spikes above the Baik--Ben~Arous--P\'ech\'e transition so that, by Proposition~\ref{prop:classic asymptotics}, each spiked sample eigenvector carries information about its population counterpart. Since the spikes are ordered, it suffices to enforce this condition solely on the smallest spike of each covariance matrix.

\subsection{Shared subspace estimation}\label{sec:rank}

In order to estimate the shared subspace, we must determine both its dimension $r_S$ and the index sets $\Psi_X$ and $\Psi_Y$, of the spiked eigenvectors that span it. Our strategy relies on a fundamental geometric property: if a direction is genuinely shared, the corresponding population eigenvectors of $\SigX$ and $\SigY$ exhibit structural alignment, whereas distinct directions do not. We can therefore detect the shared subspace by measuring the alignment between the empirical eigenspaces of the two covariance matrices. A natural framework for this task is the concept of principal angles between subspaces \citep{ye2016schubert}.

For two subspaces $\calZ, \calW \subseteq \Real^p$ of dimensions $d_Z$ and $d_W$, the principal angles $0 \le \theta_1 \le \cdots \le \theta_{\min(d_Z,d_W)} \le \pi/2$ are defined recursively by
$\cos\theta_k = \max_{z \in \calZ,\ w \in \calW} z^\top w,$
where the maximum is taken over unit vectors $z$ and $w$ that are orthogonal to all previously chosen principal vectors $z_1,\ldots,z_{k-1}$ and $w_1,\ldots,w_{k-1}$, respectively. The maximizing pair $(z_k, w_k)$ is called the $k$-th pair of principal vectors. Equivalently, and more usefully for computation, if the columns of $Z$ and $W$ form orthonormal bases of $\calZ$ and $\calW$, then the cosines of the principal angles are the singular values of $Z^\top W$, 
$\cos\theta_k = \sigma_k(Z^\top  W), \; k = 1,\ldots,\min(d_Z,d_W).$

Under the PSS model \eqref{eq:PSS}, the shared subspace is precisely the intersection of $\tspan(Q)$ and $\tspan(U)$, which implies that exactly $r_S$ of the principal angles between these two spans are zero. Equivalently, $r_S$ singular values of $Q^\top  U$ are equal to one:
\[
r_S = \sum_k \mathbb I \big(\sigma_k(Q^\top U) = 1 \big )
\]
where $\mathbb I (\cdot)$ denotes the indicator function. In the sample domain, however, this clean geometric structure is obscured because noise ensures the singular values $\hQ^\top  \hU$ are strictly less than one.  Nevertheless, under Assumptions~\ref{ass:high-dim regime}--\ref{ass:above transition}, the sample singular values converge to deterministic limits. Although the singular value sets $\sigma(\hQ_S^\top \hU_S)$ and $\sigma(\hQ_D^\top \hU_D)$ in the theorem below are oracle objects, in the sense that they require knowledge of the unobserved index sets $\Psi_X$ and $\Psi_Y$, characterizing their limiting behavior is essential, as these limits directly drive our empirical estimation procedure.

\begin{thm}\label{thm:singular convergence}
    Under Assumptions~\ref{ass:high-dim regime} and
    \ref{ass:above transition}, the ordered singular values converge almost surely, with limiting multisets
    \[
    \sigma(\hQ_S^\top  \hU_S) \aslim \sigma(\Phi_{X,S} R\, \Phi_{Y,S})
    \qquad \text{and} \qquad
    \sigma(\hQ_D^\top  \hU_D) \aslim \sigma(\Phi_{X,D} Q_D^\top  U_D\, \Phi_{Y,D})
    \]
    as $n_X, n_Y \to \infty$, where $\Phi_{X,S}$ and $\Phi_{Y,S}$ are the $r_S$-dimensional diagonal matrices with entries $\phi_{c_X}(\lambda_{i,S})$ and $\phi_{c_Y}(\gamma_{j,S})$, and $\Phi_{X,D}$ and $\Phi_{Y,D}$ are the $d_X$- and $d_Y$-dimensional diagonal matrices with entries $\phi_{c_X}(\lambda_{i,D})$ and $\phi_{c_Y}(\gamma_{j,D})$. Consequently,
    \[
    \sigma(\hQ^\top  \hU) \aslim
    \sigma(\Phi_{X,S} R\, \Phi_{Y,S}) \,\uplus\,
    \sigma(\Phi_{X,D} Q_D^\top  U_D\, \Phi_{Y,D}),
    \]
    where $\uplus$ denotes the multiset union and convergence holds element-wise in the ordered singular values. 
\end{thm}

Theorem~\ref{thm:singular convergence} decomposes the limiting singular values of $\hQ^\top  \hU$ according to their subspace of origin. The singular values of $\hQ_S^\top  \hU_S$, which quantify the alignment between the sample shared eigenspaces, converge to $\sigma(\Phi_{X,S} R\, \Phi_{Y,S})$. Conversely, those of $\hQ_D^\top  \hU_D$, which measure the alignment between the distinct eigenspaces, converge to $\sigma(\Phi_{X,D} Q_D^\top  U_D\, \Phi_{Y,D})$. Because the shared and distinct subspaces are mutually orthogonal across the two datasets, the cross-block terms $\hQ_S^\top  \hU_D$ and $\hQ_D^\top  \hU_S$ vanish asymptotically. Consequently, the matrix $[\hQ_S,\hQ_D]^\top  [\hU_S,\hU_D]$ becomes asymptotically block-diagonal, and complete spectrum of $\hQ^\top\hU$ converges to the multiset union of these two limits.

The limiting values of $\sigma(\hQ_S^\top \hU_S)$ depend on the aspect ratios $c_X, c_Y$, the shared spikes $\lambda_{i,S}, \gamma_{j,S}$, and the rotation $R$. Theorem~\ref{thm:singular convergence} suggests estimating $r_S$ by counting the singular values of $\hQ^\top \hU$ that exceed the smallest shared limit, $\min\sigma(\Phi_{X,S} R\, \Phi_{Y,S})$. For this approach to recover $r_S$, the shared singular values must be separated from the distinct ones so that a threshold isolates the shared subspace. This threshold, however, is an oracle object and practically unusable. It depends on the unknown shared spikes and the rotation $R$, and is inherently circular: since $\Phi_{X,S}, \Phi_{Y,S}$, and $R$ are $r_S$-dimensional, the threshold cannot be formed without already knowing $r_S$. Fortunately, it admits a tight lower bound that removes the dependence on $R$ and reduces the remaining dependence to a single eigenvalue from each dataset.

\begin{lem}\label{lem:rotation inequality}

For any rotation matrix $R$, 
\[
\min\big(\sigma(\Phi_{X,S} R\, \Phi_{Y,S})\big)
\;\ge\;
\phi_{c_X}(\lambda_{r_S,S})\,\phi_{c_Y}(\gamma_{r_S,S}),
\]
with equality when $R = I_{r_S}$.
\end{lem}

This lower bound, free of $R$ and depending only on the smallest shared spikes, serves as our cutoff value. For it to recover $r_S$, the distinct singular values must fall below the bound while the shared ones lie above, which requires the alignment between sample distinct subspaces to be sufficiently separated from the one between sample shared subspaces. We now assume exactly this.

\begin{assumption} \label{ass:shared distinct separable}
    The distinct subspaces of $\SigX$ and $\SigY$ are separated from the shared subspace in the sense that 
    \[
    \phi_{c_X}(\lambda_{r_S,S}) \phi_{c_Y}(\gamma_{r_S,S}) > \sigma_{1}\big(\Phi_{X,D}Q_D^\top U_D\Phi_{Y,D}\big).
    \]
\end{assumption}

This assumption naturally holds in high dimensions, where independent, distinct subspaces are expected to be nearly orthogonal. Indeed, if they are mutually orthogonal, the right-hand side vanishes and the condition holds trivially. Conversely, the assumption could fail under two scenarios: first, if a distinct direction exhibits strong cross-dataset alignment while carrying a large spike, in which case it behaves more like a shared component than a distinct one, or second, if a genuinely shared direction has a signal too weak to be distinguished from the distinct noise bulk.

Although free of $R$, the cutoff still depends on the true eigenvalues $\lambda_{r_S,S}$ and $\gamma_{r_S,S}$. A natural remedy is to substitute their sample counterparts. However, sample spiked eigenvalues are asymptotically biased upward, which rules out direct plug-in estimation. We resolve this by inverting the asymptotic bias mapping from Proposition~\ref{prop:classic asymptotics} using the debiasing function
\[
d(\ell, c) = \frac{(\ell + 1 - c) + \sqrt{(\ell + 1 - c)^2 - 4\ell}}{2},
\qquad \ell > (1 + \sqrt{c})^2. 
\]
When applied to an over-threshold sample eigenvalue, $d(\ell, c)$ yields a consistent estimate of its population counterpart. Letting $\tilde\lambda_{r_S,S} = d(\hat\lambda_{r_S,S}, c_X)$ and $\tilde\gamma_{r_S,S} = d(\hat\gamma_{r_S,S}, c_Y)$ denote the debiased minimal shared spikes, we can use them to form our threshold. However, because the index sets $\Psi_X$ and $\Psi_Y$ are unobserved, identifying the minimal shared sample spikes $\hat\lambda_{r_S,S}$ and $\hat\gamma_{r_S,S}$  requires knowledge of $\max(\Psi_X)$ and $\max(\Psi_Y)$. Thus, we define the resulting oracle estimator for $r_S$ as
\begin{equation}\label{eq:rank}
\hrS^* = \sum_{k} \mathbb{I}\big(\sigma_k(\hQ^\top \hU) \ge \phi_{c_X}(\tilde\lambda_{r_S,S})\,\phi_{c_Y}(\tilde\gamma_{r_S,S})\big),
\end{equation}
where the asterisk explicitly denotes this oracle dependence. The following theorem shows that the oracle estimator $\hrS^*$ is consistent for $r_S$. 

\begin{thm}\label{thm:rank consistency}
Under Assumptions~\ref{ass:high-dim regime}, \ref{ass:above transition}, and \ref{ass:shared distinct separable},
\[
\hrS^* \aslim r_S \qquad \text{as}\quad  n_X, n_Y \to \infty.
\]
\end{thm}

We now address the estimation of the shared index sets $\Psi_X$ and $\Psi_Y$ under the assumption that $r_S$ is known. Let $\hP_X = \hQ\hQ^\top$ and $\hP_Y = \hU\hU^\top$ denote the projection operators onto the column spaces of the sample eigenvector matrices $\hQ$ and $\hU$, respectively. Similarly, let $\hP_{X,i} = \hq_i \hq_i^\top$ and $\hP_{Y,j} = \hu_j \hu_j^\top$ denote the rank-one projections onto the individual eigenvectors for $i = 1,\ldots,r_X$ and $j = 1,\ldots,r_Y$. For a symmetric positive semi-definite matrix $A$, let $\Pi_r(A)$ denote the projection operator onto its leading $r$ eigenvectors. Then the matrix $\bar P_{r_S} = \Pi_{r_S}(\hP_X + \hP_Y)$ serves as an estimate of the projection onto the shared subspace. By pooling the operators, $\hP_X + \hP_Y$ accumulates signal across both datasets, ensuring its leading $r_S$ eigenvectors isolate their common directions. The shared eigenvectors within each dataset are those whose individual projection operators align most closely with $\bar P_{r_S}$, motivating the estimators
\begin{equation}\label{eq:index}
\hPsiX^* = \argmax_{i;\, r_S} \big\| \bar P_{r_S}\, \hP_{X,i} \big\|_F,
\qquad
\hPsiY^* = \argmax_{j;\, r_S} \big\| \bar P_{r_S}\, \hP_{Y,j} \big\|_F,
\end{equation}
where $\argmax_{i;\,r_S}$ returns the set of $r_S$ indices corresponding to the largest values of the objective function. 

The following theorem states that this construction recovers the shared index sets. Specifically, given the true rank $r_S$, the oracle estimators $\hPsiX^*$ and $\hPsiY^*$ are consistent.

\begin{thm} \label{thm:index consistency}
Under Assumptions \ref{ass:high-dim regime}, \ref{ass:above transition}, and \ref{ass:shared distinct separable},
\[
\mathbb P\big(\hPsiX^* = \Psi_X  \text{ and } \hPsiY^* = \Psi_Y \text{ for all sufficiently large } n_X, n_Y\big) = 1.
\]
\end{thm}

The estimation in \eqref{eq:rank} and \eqref{eq:index} presents a circular dependence. The first estimates $r_S$ consistently, but only given $\max(\Psi_X)$ and $\max(\Psi_Y)$; the second estimates the index sets $\Psi_X$ and $\Psi_Y$ consistently, but only given $r_S$. Neither oracle quantity can be formed without the output of the other. We propose an iterative method to resolve this, alternating between the two estimation steps. It is initialized with $\hPsiX = \{r_X\}$ and $\hPsiY = \{r_Y\}$, the largest indices of each dataset, so that no shared direction is missed at the onset. Each subsequent iteration updates $\hrS$ given the current index sets, then updates the index sets given the new $\hrS$, stopping once the index sets no longer change. We present Algorithm~\ref{alg:shared rank}, which formalizes this estimation scheme, in Appendix~\ref{sec:appendix-algorithm}. 

Note that while Theorems~\ref{thm:rank consistency} and \ref{thm:index consistency} establish consistency of the two oracle estimators, they do not automatically guarantee the consistency of the empirical estimator $\hrS$ output by Algorithm~\ref{alg:shared rank}. Furthermore, algorithmic termination is not universally guaranteed, because the algorithm defines a deterministic map over a finite state space, its trajectory must eventually either converge to a fixed point or enter a periodic cycle. Although non-convergence can be structurally prevented by enforcing a maximum iteration threshold, in practice, the algorithm demonstrates exceptional convergence properties: across all replications in our simulation study (Section~\ref{sec:simulation}), a stable fixed point is reached in only a few iterations.

\subsection{Target covariance estimation}\label{sec:covariance}

We now estimate the target covariance matrix $\Sigma_X$ using the shared rank $r_S$ and the index sets $\Psi_X$ and $\Psi_Y$. Our proposed covariance estimator is built upon the estimate of the projection matrix onto the true shared subspace, 
\[
P_S = \QS \QS^\top  = \US \US^\top.
\]
Once $P_S$ is identified, the remaining component of $\Sigma_X$ is obtained by projecting the spiked part of $\Sigma_X$ onto the complement of the shared subspace. Concretely, the spiked part of $\Sigma_X$ is decomposed as
\begin{equation}\label{eq:Sigma_X decomp}
\Sigma_X - I_p
= P_S (\Sigma_X - I_p) P_S
+ (I_p - P_S)(\Sigma_X - I_p)(I_p - P_S),
\end{equation}
where the first term is the shared component with rank $r_S$ and the second is the distinct component with rank $d_X$. Plugging an estimator $\hat P_S$ and the sample spiked component estimate $\widehat{\Sigma_X - I_p}:=\hat Q \hat\Lambda \hat Q^\top$  yields the plug-in estimator
\begin{equation}\label{eq:Sigma_X estimator}
\hat\Sigma_X
= \hat P_S\, \hat Q \hat\Lambda \hat Q^\top \, \hat P_S
+ \mathcal{T}_{d_X}\big( (I_p - \hat P_S)\, \hat Q \hat\Lambda \hat Q^\top \, (I_p - \hat P_S)\big)
+ I_p,
\end{equation}
where $\mathcal{T}_r(A)$ is the rank $r$ truncation operator for symmetric positive semi-definite matrix $A$. 

In what follows, we discuss how to combine the information from both datasets to estimate $P_S$. A natural starting point pools the total sample spiked projections $\hat P_X = \hQ\hQ^\top $ and $\hat P_Y = \hU\hU^\top $ and extracts their leading $r_S$ directions, i.e., $\bar P_{r_S} = \Pi_{r_S}\big(\hat P_X + \hat P_Y\big),$ motivated by the fact
\[
P_S = \Pi_{r_S}\big(QQ^\top  + UU^\top \big) = \Pi_{r_S}\big(Q_S Q_S^\top  + U_S U_S^\top \big).
\]
We propose to utilize $\hat P_X$ and $\hat P_Y$ rather than $\hQ_S\hQ_S^\top $ and $\hU_S\hU_S^\top $, since the latter would induce additional uncertainty as it depends on the estimated index sets $\hat\Psi_X, \hat\Psi_Y$ in practice. Furthermore, $\hQ_S$ and $\hU_S$ exhibit instability due to perturbation from adjacent distinct spiked directions even if we know the true index sets. A simulation study comparing the two approaches is reported in the Appendix. 

Our final estimator of the shared projection matrix $P_S$ applies more general pooling with possibly unequal weights for target and background, in order to reflect the possibly different signal strengths of the two datasets: 
\[
\hat P_S^{(\alpha)} = \Pi_{r_S}\big(\alpha\, \hat P_X + (1-\alpha)\, \hat P_Y\big),
\qquad \alpha \in (0,1). 
\]
We propose to choose $\alpha$ so that our estimator minimizes the asymptotic Frobenius loss, whose approximation is presented in the next theorem.

\begin{thm} \label{thm:Frobenius performance}
Under Assumptions~\ref{ass:high-dim regime}, \ref{ass:above transition}, and \ref{ass:shared distinct separable}, for $\alpha \in (0,1)$,
\[
\tfrac{1}{2}\big\| P_S - \hat P_S^{(\alpha)} \big\|_F^2 \aslim L(\alpha)
\qquad \text{as } n_X, n_Y \to \infty,
\]
where $L(\alpha)$ is given in Appendix \ref{sec:appendix-proof}. The asymptotic loss can be approximated by
\begin{equation}\label{eq:approx performance}
L(\alpha) \approx L^\dagger (\alpha) = \alpha\, L_X + (1-\alpha)\, L_Y - \alpha(1-\alpha)\,(r_S - T),
\end{equation}
where
\[
L_X = r_S - \sum_{i=1}^{r_S}\phi_{c_X}^2(\lambda_{i,S}),
\;
L_Y = r_S - \sum_{j=1}^{r_S}\phi_{c_Y}^2(\gamma_{j,S}),
\; \text{and }\;
T = \mathrm{tr}\big(\Phi_{X,S}^2 R\, \Phi_{Y,S}^2 R^\top \big). 
\] 
\end{thm}
While the detail of this approximation is provided in Appendix \ref{sec:appendix-proof}, we note that the approximate asymptotic loss becomes more accurate when the shared singular values $\sigma_k(\Phi_{X,S} R\, \Phi_{Y,S})$ are closer to one, that is, when the two sample shared subspaces are strongly aligned, which is the regime where pooling is most useful. The first two terms are a convex combination of the asymptotic losses $L_X$ and $L_Y$ of the two single-dataset estimators: by Proposition~\ref{prop:classic asymptotics}, $\tfrac{1}{2}\| P_S - \hQ_S\hQ_S^\top \|_F^2 \aslim L_X$, and analogously for $L_Y$. The third term is the \emph{pooling gain}: $r_S - T \ge 0$, with equality only in the degenerate classical regime where every $\phi_{c_X}(\lambda_{i,S})$ and $\phi_{c_Y}(\gamma_{j,S})$ equals one --- that is, when sample eigenvectors are asymptotically consistent with their population counterparts. In any high-dimensional setting, $r_S - T > 0$ and the pooling gain $\alpha(1-\alpha)(r_S - T)$ is strictly positive. 

The optimal weight that minimizes the approximate asymptotic loss  \eqref{eq:approx performance} of the shared subspace estimation is 
\[
\alpha^* = \frac{1}{2} + \frac{L_Y - L_X}{2\,(r_S - T)}, 
\]
which governed by the difference $L_Y - L_X$ between the single-dataset losses. As each loss depends on both the aspect ratio and the shared-spike magnitudes through $\phi_{c_X}$ and $\phi_{c_Y}$, the optimal weight reflects signal strength as well as sample size so that a target with a stronger shared signal would be up-weighted even when it had fewer samples. With $\alpha = \alpha^*$, the approximate asymptotic loss becomes 
\[
L^\dagger (\alpha^*) = L_X -
b = L_X - \frac{\big((r_S - T) + (L_X - L_Y)\big)^2}{4(r_S - T)},
\]
whose second term quantifies the asymptotic gain over the single data loss $L_X$ through pooling.

The optimal weight $\alpha^*$ can be estimated by plugging in the debiased sample spikes $\tlamb_{i,S}$ and $\tgamm_{j,S}$ and using the approximation 
$
T \approx \sum_{i=1}^{r_S} \phi_{c_X}^2(\lambda_{i,S})\,
                         \phi_{c_Y}^2(\gamma_{i,S}). 
$
Denoting the estimated optimal weight as $\hat\alpha$, we adopt $\hat P_S = \hat P_S^{(\hat\alpha)}$ as our estimator of $P_S$. See Appendix~\ref{sec:appendix-pooling weight} for a simulation study on the effectiveness of this choice of the weight.

The distinct component of $\hat{\Sigma}_X$ in \eqref{eq:Sigma_X estimator} is the rank $d_X$ truncation of matrix $(I_p - \hat P_S)\,\hat Q\hat\Lambda\hat Q^\top \,(I_p - \hat P_S)$. Its eigendecomposition yields a basis for the distinct subspace, denoted by $\hat Q_D$, which can be used for a contrastive analysis, as illustrated in Section~\ref{sec:real data analysis}.

\section{Simulation study}\label{sec:simulation}

We evaluate the proposed method in two stages of simulation. The first set of experiments tests the method on data generated from the PSS model, examining how accurately it recovers the shared subspace rank $r_S$ and the target covariance matrix $\Sigma_X$ across a range of model configurations. The second set evaluates robustness by both relaxing the model's structural assumptions and restricting to special cases that match the assumptions of competing methods. 

Three existing methods are considered: The contrastive dimension estimator (CDE) \citep{hawke2024contrastive}, designed to estimate $d_X$, is used for estimating the shared rank as $\hat r_S = r_X - \hat d_X$, where $r_X$ is the true rank. This method does not provide target covariance estimation, for which we omit it from comparison. The multi-group covariance estimator (MgCov) \citep{franks2019shared} and the personalized PCA (PerPCA) \citep{shi2024personalized} provide the estimates for both $r_S$ and $\Sigma_X$.   

\subsection{Estimation under the PSS model}\label{subsec:est-sim}

We generate target and background datasets from the PSS model \eqref{eq:PSS} with shared rank $r_S = 5$ and distinct ranks $d_X = 4$ and $d_Y = 8$, giving total spiked ranks $r_X = 9$ and $r_Y = 12$. The spiked eigenvalues of $\Sigma_X$ are set to $\lambda_i = e^{\ell_i}$ with $\ell_1 > \cdots > \ell_{r_X}$ equally spaced between $3.5$ and $1$, and the spiked eigenvalues of $\Sigma_Y$ are set to $\gamma_j = e^{g_j}$ with $g_1 > \cdots > g_{r_Y}$ equally spaced between $4$ and $1$, so that the background spikes are slightly larger overall. We sample $Q_S \in \calO^{p\times r_S}$ uniformly from the Stiefel manifold and set $U_S = Q_S R$ for a uniformly sampled $r_S \times r_S$ rotation matrix $R$.

We vary two aspects of the configuration. The first aspect regards the relationship between the distinct eigenvector matrices $Q_D$ and $U_D$:
\begin{enumerate}[label=(\roman*)]
  \item \emph{Orthogonal}: $Q_D^\top U_D = 0$;
  \item \emph{Non-orthogonal}: the principal angles between
        $\mathrm{span}(Q_D)$ and $\mathrm{span}(U_D)$ are equally spaced
        between $0.1$ and $0.3$.
\end{enumerate}
The second aspect is about the positioning of the shared spikes in the spectrum, encoded by the index sets $\Psi_X$ and $\Psi_Y$:
\begin{enumerate}[label=(\alph*)]
  \item \emph{Top}: $\Psi_X = \Psi_Y = \{1, 2, 3, 4, 5\}$, i.e., the shared
        directions are the leading spikes of both datasets;
  \item \emph{Mixed}: $\Psi_X = \{2, 4, 5, 6, 7\}$ and $\Psi_Y = \{3, 4, 5, 8, 11\}$, i.e., the shared directions are interleaved with the distinct ones.
\end{enumerate}

For each combination of (i)/(ii) and (a)/(b), we sample $X_1, \ldots, X_{n_X} \stackrel{\text{i.i.d.}}{\sim} N_p(0, \Sigma_X)$ and $Y_1, \ldots, Y_{n_Y} \stackrel{\text{i.i.d.}}{\sim} N_p(0, \Sigma_Y)$, in two dimensionality regimes: $(p, n_X, n_Y) = (200, 50, 200)$ and $(1000, 250, 1000)$, both with $p/n_X = 4$ and $p/n_Y = 1$. Each setting is replicated $100$ times.


We first report the results on the shared rank estimation in Table~\ref{table:simul1}. The proposed PSS method produces the most accurate estimates. Though it shows difficulty in the small-sample, mixed-position settings, other methods deteriorate even further. The CDE method significantly underestimates $r_S$ in all settings, while MgCov and PerPCA consistently overestimate it. It is also noticeable that CDE and MgCov perform worse as the dimension and sample size increase, especially in the mixed-position setting. This contrasts with the fact that PSS estimates concentrate around the true value $r_S = 5$ and become less variable, as expected from the asymptotics in Section~\ref{sec:estimation}. Furthermore, PSS still maintains relatively strong performance in both the mixed-position and non-orthogonal scenarios.

The rightmost column of Table~\ref{table:simul1} reports the number of correctly identified eigenvector indices by PSS for the shared subspace, i.e., $|\hat\Psi_X \cap \Psi_X|$. Note that none of the competing methods returns index sets. In all cases except the two with small samples and mixed indices, the PSS identifies the correct set with high accuracy.

\begin{table}[t]
\centering
\resizebox{\linewidth}{!}{%
\begin{tabular}{lll cccc c}
\hline
\multirow{2}{*}{\shortstack{Position of\\$\Psi_X, \Psi_Y$}}
  & \multirow{2}{*}{$\angle(Q_D, U_D)$}
  & \multirow{2}{*}{$(p, n_X, n_Y)$}
  & \multicolumn{4}{c}{$\hat r_S$}
  & $|\hat\Psi_X \cap \Psi_X|$ \\
\cline{4-8}
  &  &  & PSS & CDE & MgCov & PerPCA
  & PSS \\
\hline
\multirow{4}{*}{(a) Top}
  & \multirow{2}{*}{(i) Orthogonal}
    & $(200, 50, 200)$   & 4.58 (0.05) & 1.93 (0.06) & 9.96 (0.03) & 13.1 (0.25) & 4.55 (0.05) \\
  & & $(1000, 250, 1000)$ & 4.85 (0.03) & 1.86 (0.04) & 10.5 (0.05) & 12.3 (0.33) & 4.84 (0.03) \\
\cline{2-8}
  & \multirow{2}{*}{(ii) Non-orthogonal}
    & $(200, 50, 200)$    & 4.65 (0.05) & 1.92 (0.05) & 10.0 (0.05) & 13.6 (0.23) & 4.61 (0.05) \\
  & & $(1000, 250, 1000)$ & 4.89 (0.03) & 1.77 (0.04) & 10.6 (0.05) & 13.4 (0.25) & 4.89 (0.03) \\
\hline
\multirow{4}{*}{(b) Mixed}
  & \multirow{2}{*}{(i) Orthogonal}
    & $(200, 50, 200)$    & 3.02 (0.05) & 0.12 (0.03) & 12.1 (0.06) & 13.6 (0.22) & 2.43 (0.07) \\
  & & $(1000, 250, 1000)$ & 4.96 (0.01) & 0.03 (0.02) & 13.0 (0.01) & 11.7 (0.34) & 4.95 (0.02) \\
\cline{2-8}
  & \multirow{2}{*}{(ii) Non-orthogonal}
    & $(200, 50, 200)$    & 3.07 (0.05) & 0.74 (0.04) & 11.8 (0.06) & 14.0 (0.21) & 2.51 (0.07) \\
  & & $(1000, 250, 1000)$ & 4.93 (0.02) & 0.36 (0.05) & 12.8 (0.04) & 13.6 (0.34) & 4.88 (0.03) \\
\hline
\end{tabular}
}
\caption{Simulation results for the shared rank and the corresponding index estimation under the PSS model. Standard errors in parentheses.}
\label{table:simul1}
\end{table}


\begin{figure}[!h]
    \centering
    \includegraphics[width=1\linewidth]{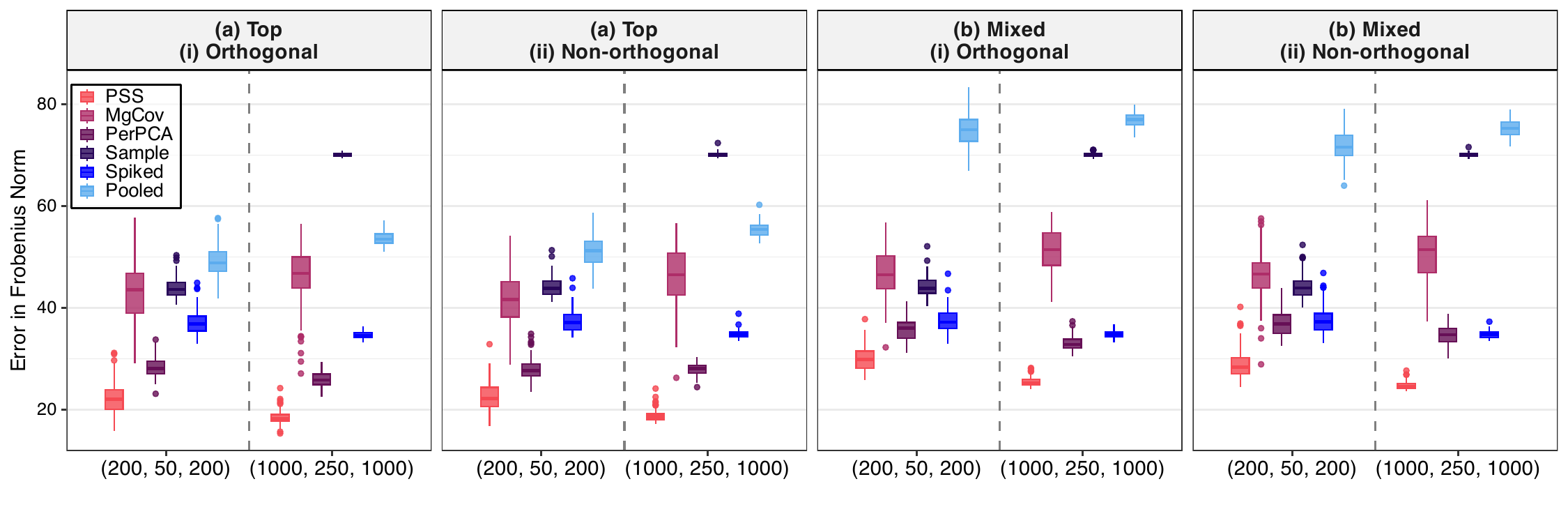}
    \caption{Frobenius error $\|\hat\Sigma_X - \Sigma_X\|_F$ for target
covariance estimation under the PSS model, across the four configurations
(shared-index position $\times$ orthogonality of distinct subspaces) and
two $(p,n_X,n_Y)$ regimes. Boxplots over $100$ replications.}
    \label{fig:sim estcov exact model}
\end{figure}

We next assess the accuracy of the target covariance estimate, measured by the Frobenius error $\|\hat\Sigma_X - \Sigma_X\|_F$. In addition to the three competing methods, we include three different estimates of the target covariance: the sample covariance matrix $S_X$, the spiked covariance estimate $\hat Q\hat\Lambda\hat Q^\top + I_p$, and the na\"ively pooled sample covariance $(n_X + n_Y)^{-1}(n_X S_X + n_Y S_Y)$.

The boxplots in Figure~\ref{fig:sim estcov exact model} show that PSS achieves the lowest error across all configurations. Its accuracy improves as dimensionality and sample size grow in proportion, while the competing methods and the sample and pooled baselines deteriorate; only the spiked estimator improves at the same rate as PSS. The robustness pattern of the rank-estimation results carries over. PSS remains accurate under the mixed-position scenario~(b) and the non-orthogonal scenario~(ii), while PerPCA, MgCov, and the spiked baseline deteriorate sharply in those configurations.

\subsection{Model misspecification scenarios} \label{subsec: sim beyond pcs}

\begin{figure}[t]
    \centering
    \includegraphics[width=1\linewidth]{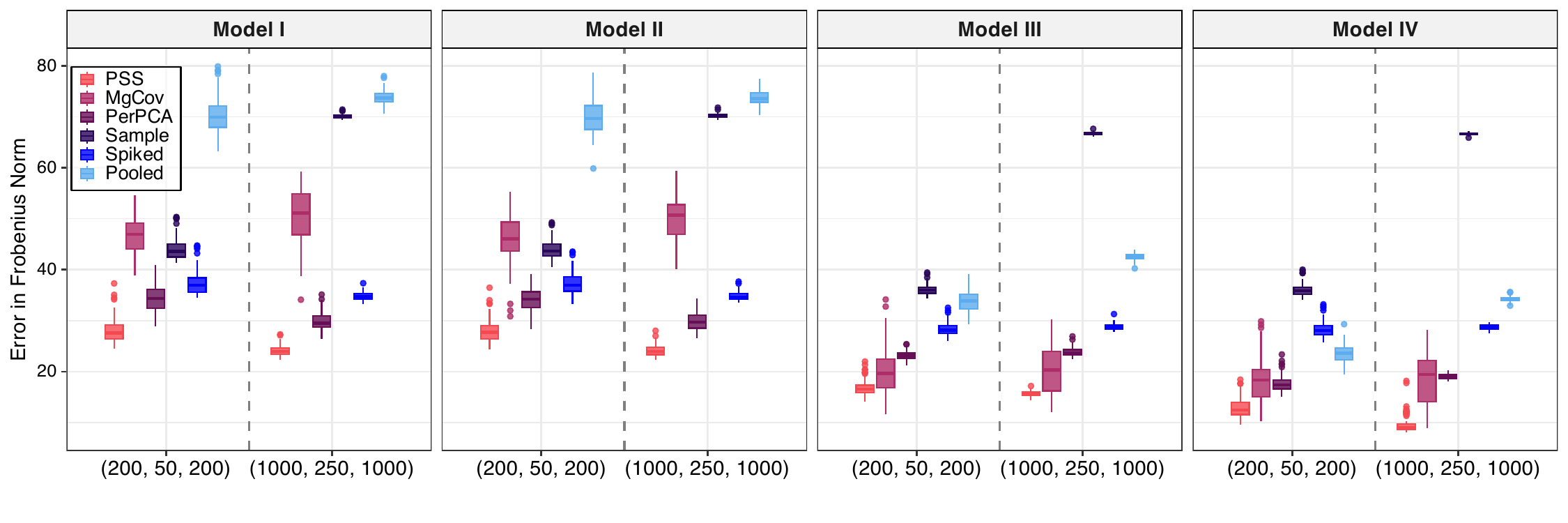}
    \caption{Estimation errors for target covariance estimation under four departures from the PSS model: Model~I (misspecification, shared subspaces approximately but not exactly aligned), Models~II--IV (increasingly restricted special cases).}
    \label{fig:sim estcov missed model}
\end{figure}

We now examine how PSS behaves when the model assumptions are violated or when the underlying covariance structure is restrictive, rendering the flexibility of the PSS model unnecessary. We consider four models: Model I violates the PSS assumption of exactly aligned shared subspaces: $\mathrm{col}(Q_S) \ne \mathrm{col}(U_S)$ but $\mathrm{col}(Q_S) \approx \mathrm{col}(U_S)$ with principal angles of $5^\circ$, with distinct components present. Models II--IV are special cases with increasingly restrictive structure: Model II sets $Q_S = U_S$, with distinct components present; Model III has $\mathrm{col}(Q_S) = \mathrm{col}(U_S)$ with no distinct components ($d_X = d_Y = 0$); and Model IV sets $Q_S = U_S$ with no distinct components. Note that Model II--IV match the assumptions of MgCov or PerPCA (see Remark \ref{rmk:franks}). The exact data-generating processes are detailed in Appendix~\ref{appendix: data gen}. Each model is replicated $100$ times, and the resulting estimation errors are shown in Figure~\ref{fig:sim estcov missed model}. For Models~I and II, the overall pattern of the boxplots is similar to that of Figure~\ref{fig:sim estcov exact model}. PSS substantially outperforms all competing methods across all settings. MgCov is a close second in Models III and IV, which are exactly the underlying structure it targets. However, it has a noticeably larger variance than other methods. 

Overall, the simulation results suggest that the proposed method is broadly applicable: it dominates the compared methods under the general PSS model \ref{eq:PSS} as seen in Section \ref{subsec:est-sim}, handles more restrictive cases (Models II--IV), and remains robust for a misspecified case (Model I). Additional simulation studies on degenerate cases are given in Appendix~\ref{sec:appendix-degenerate PSS}.

\section{Real data applications}\label{sec:real data analysis}

We demonstrate the utility of PSS with two real-world datasets with a natural target--background structure. In Section~\ref{sec:sp500}, we use the proposed method to construct a minimum-variance portfolio during the early COVID-19 pandemic using pre-pandemic returns as background data. In Section~\ref{sec:lgg}, we characterize tumor-specific signals in gene expression of low-grade glioma patients using normal brain tissues as background. 

\subsection{Minimum-variance portfolio during pandemic}
\label{sec:sp500}

A global minimum-variance (GMV) portfolio is a benchmark of modern portfolio theory: given an estimated covariance matrix $\hat\Sigma$ of $p$ assets, the GMV portfolio weights are $\hat w_{\text{GMV}} = (\boldsymbol{1}^\top \hat\Sigma^{-1}\boldsymbol{1})^{-1}\hat\Sigma^{-1}\boldsymbol{1}$, where $\boldsymbol{1}$ is the $p$-vector of ones. Although the GMV portfolio ignores expected returns, it is widely used as a risk-minimizing baseline and often outperforms alternatives in out-of-sample tests \citep{jagannathan2003risk}. Its quality depends entirely on $\hat\Sigma$, making covariance estimation the central problem.

We apply PSS to GMV portfolio construction at the onset of the COVID-19 pandemic, when accurate covariance estimation is especially difficult. Stock market dynamics shifted substantially during this period \citep{horvath2021examination, wu2022analysis}: estimators that use only the early-pandemic period face severe sample-size limitations, while estimators that indiscriminately pool pandemic and pre-pandemic data suffer from structural mismatch. PSS's target--background framing fits this setting naturally, with early-pandemic returns as the target and pre-pandemic returns as background. We use the constituents of the S\&P 500 index on February 28, 2020, with daily log-returns over the early-pandemic period (December 2019--February 2020) as the target and over the preceding 18 months as background. After preprocessing, $p = 499$, $n_X = 61$, and $n_Y = 378$. Based on $\hat r_X = 10$ and $\hat r_Y = 35$ estimates, we obtain $\hat r_S = 9$. The out-of-sample standard deviation of the GMV portfolio return is evaluated over March 2020.

\begin{figure}[h]
    \centering
    \includegraphics[width=0.9\linewidth]{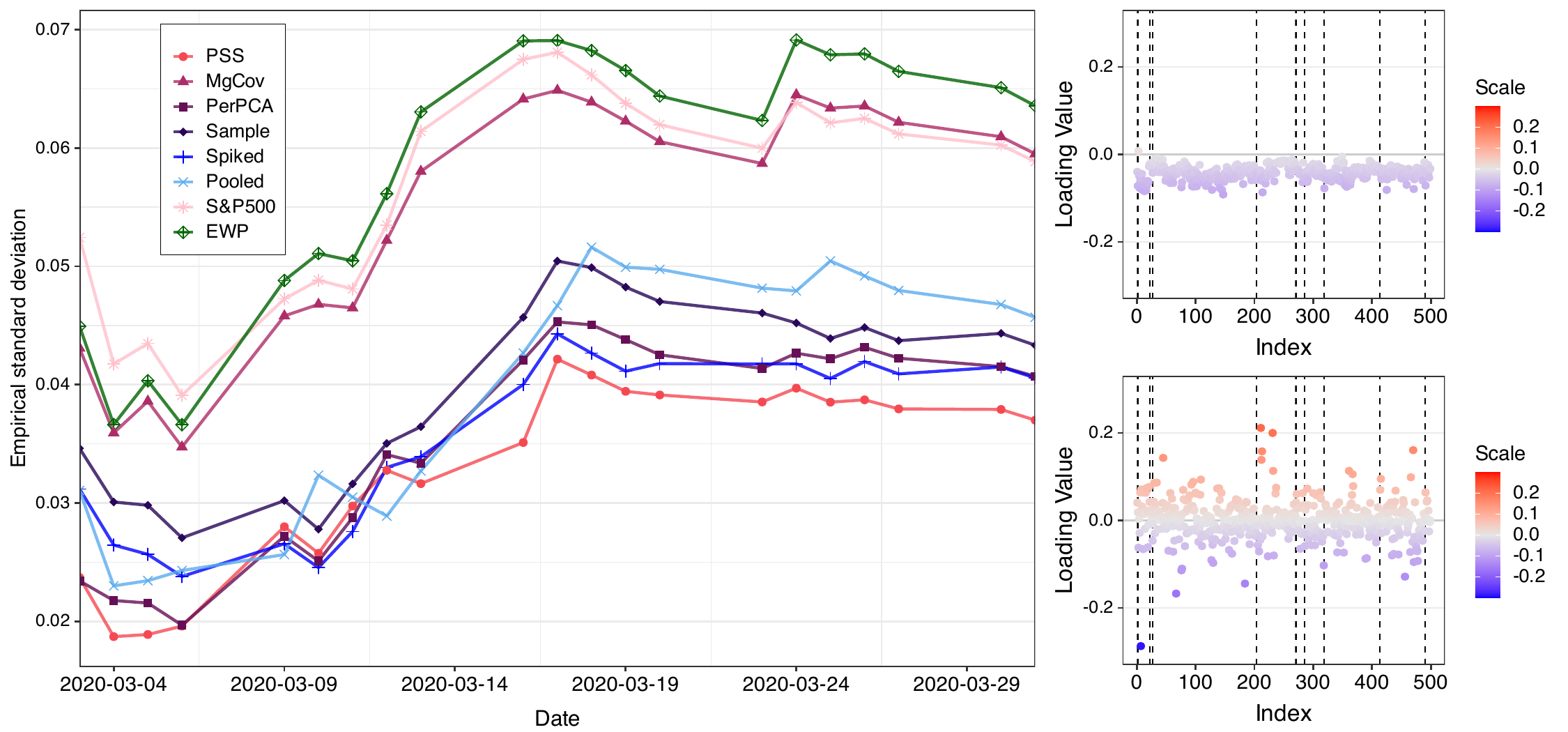}
    \caption{(Left) Out-of-sample standard deviation of GMV portfolio returns over March 2020. The S\&P 500 index and the equal-weighted portfolio (EWP) are included as non-optimized benchmarks. (Top-right) Loadings of the first shared eigenvector $\hat q_{S,1}$. (Bottom-right) Loadings of the first distinct eigenvector $\hat q_{D,1}$.}
    \label{fig:GMVP}
\end{figure}

The left panel in Figure~\ref{fig:GMVP} compares the out-of-sample standard deviations of portfolio returns over March 2020. The PSS portfolio achieves the lowest volatility throughout the validation period. MgCov produces a portfolio whose volatility tracks the S\&P 500 index and the equal-weighted portfolio, indicating that optimization based on the MgCov estimator yields no advantage over an unoptimized benchmark. PerPCA tracks the spiked target estimator closely, suggesting that the background data is not effectively leveraged.

We also investigate the shared and distinct signals identified by the proposed method. The right panels in Figure~\ref{fig:GMVP} show the first eigenvector of the shared and distinct (target) subspaces, respectively, with vertical dashed lines indicating the ten standard industrial sectors. The shared eigenvector $\hat q_{S,1}$ represents a consistent systemic risk factor that reflects overall market movement, which is the dominant factor in classical asset pricing \citep{elton2009modern}. The loadings of the distinct eigenvector $\hat q_{D,1}$ have several large absolute values. Pandemic-vulnerable businesses such as apparel retail (PVH, ROST), payments (V), cruise lines (RCL), construction machinery (DE), and oil and gas (DVN) have large positive values. On the other hand, pandemic beneficiaries such as technology and communications (TXN, CMCSA, APH), insurance analytics (VRSK), and mask manufacturing (MMM) have large negative loadings.

\subsection{Contrastive analysis of LGG gene expression}\label{sec:lgg}

Low-grade glioma (LGG) is a slow-growing brain tumor that requires long-term clinical follow-up \citep{karsonovich2025low}. It comprises two molecular subtypes, astrocytoma and oligodendroglioma, which differ in expression profile. We use RNA-seq data from $509$ LGG tumor samples in the Cancer Genome Atlas (TCGA) as the target dataset and $1{,}152$ normal brain tissue samples from the Genotype-Tissue Expression (GTEx) database as background, both obtained from the UCSC Xena platform \citep{goldman2020visualizing}. The original 60{,}498 genes are reduced to $p = 1{,}000$ by standard variance filtering.

\begin{figure}[h]
\centering
\begin{subfigure}{.19\textwidth}
  \centering
  \includegraphics[width=\linewidth]{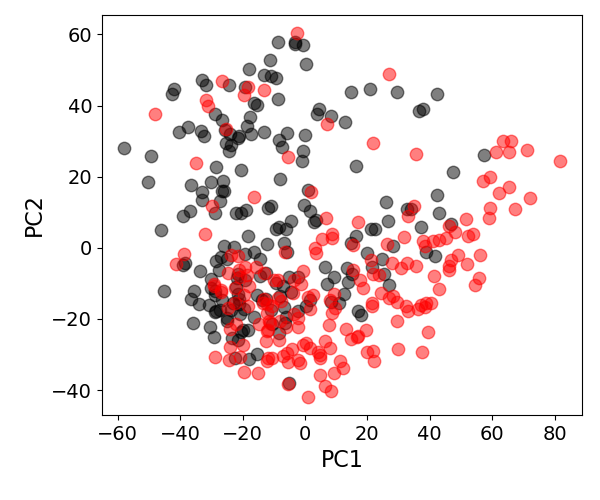}
  \caption{PCA}
\end{subfigure}
\begin{subfigure}{.19\textwidth}
  \centering
  \includegraphics[width=\linewidth]{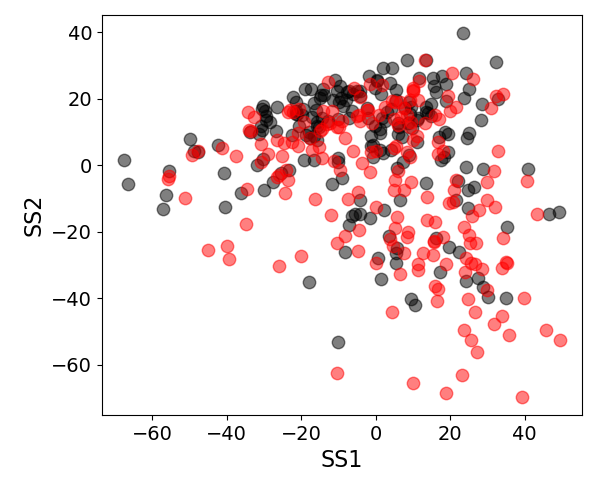}
  \caption{PSS-shared}
\end{subfigure}
\begin{subfigure}{.19\textwidth}
  \centering
  \includegraphics[width=\linewidth]{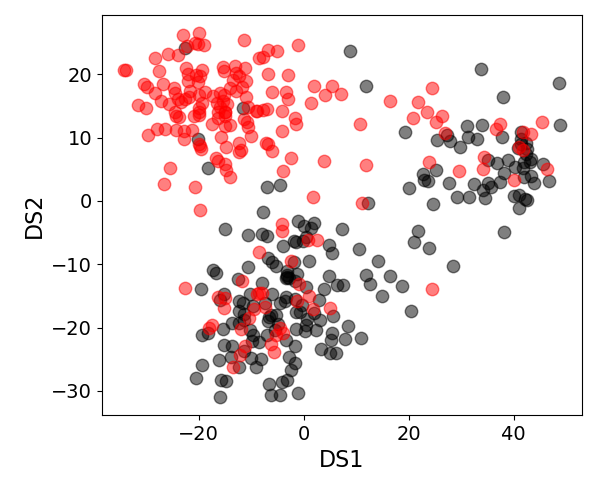}
  \caption{PSS-distinct}
\end{subfigure}
\begin{subfigure}{.19\textwidth}
  \centering
  \includegraphics[width=\linewidth]{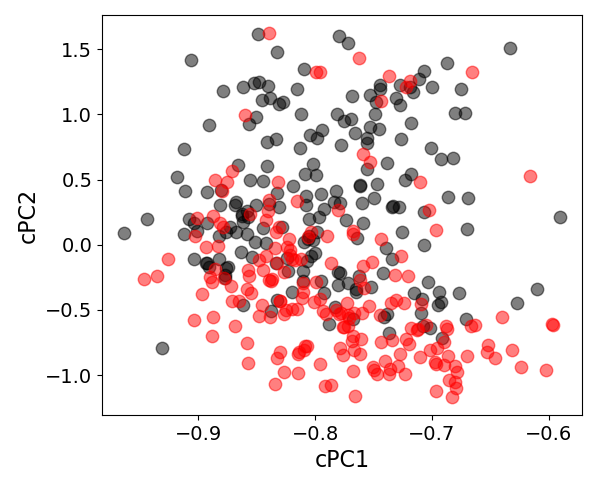}
  \caption{pcPCA}
\end{subfigure}
\begin{subfigure}{.19\textwidth}
  \centering
  \includegraphics[width=\linewidth]{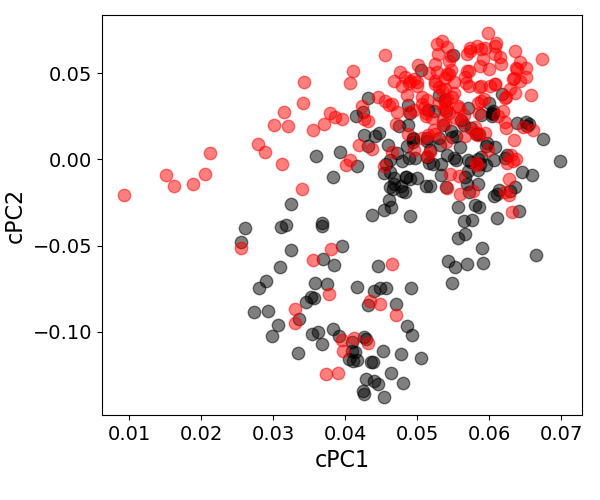}
  \caption{gcPCA}
\end{subfigure}
\caption{LGG samples projected onto: (a) Eigenspace from PCA with tumor data, (b) the shared and (c) the distinct subspaces estimated by PSS with normal brain tissue as background, and the contrastive subspaces from (d) pcPCA and (e) gcPCA. Astrocytoma and oligodendroglioma subtypes are colored as black and red, respectively.}
\label{fig:LGG projections}
\end{figure}


Figure~\ref{fig:LGG projections} provides 2-d views of LGG samples projected onto five subspaces: the full eigenspace from PCA on the tumor data, the shared and distinct subspaces from PSS, and the contrastive subspaces from the probabilistic contrastive PCA (pcPCA) by \citet{li2024probabilistic} and the generalized contrastive PCA (gcPCA) by \citet{de2025identifying}. The PSS distinct subspace cleanly separates the two tumor subtypes, while the shared subspace, which retains only the structure common to tumor and normal tissue, does not, as expected. Both pcPCA (run with the contrast parameter $\gamma = 0.2$, the only value for which the implementation converged) and gcPCA fail to recover the subtype separation. This illustrates that the proposed PSS method captures a distinct subspace containing target-specific variation that ordinary PCA misses and that existing CDR methods fail to recover in the high-dimensional regime.

\section{Discussion}\label{sec:discussion}

We proposed a general high-dimensional framework that leverages background data to estimate the covariance matrix of the target data. We considered a spiked covariance model with a partially shared subspace and proposed a complete estimation procedure, including an iterative algorithm for estimating shared rank and index sets, a plug-in covariance estimator with a closed-form optimal pooling weight, and asymptotic theory in the proportional-growth regime. 

A practical refinement concerns the finite-sample behavior of the cutoff in Algorithm~\ref{alg:shared rank}. The cutoff $\phi_{c_X}(\tlamb_{r_S,S})\phi_{c_Y}(\tgamm_{r_S,S})$ aims an asymptotic value, and in finite samples the smallest shared singular value $\sigma_{r_S}(\hQ_S^\top \hU_S)$ fluctuates around its limit and can fall below it, causing the shared rank to be underestimated. Relaxing the cutoff to $\phi_{c_X}(\tlamb_{r_S,S})\phi_{c_Y}(\tgamm_{r_S,S}) - \epsilon$ for a small nonnegative margin $\epsilon$ guards against this at the risk of admitting spurious shared directions. Appendix~\ref{sec:epsilon} compares several choices and finds that none is uniformly best; a principled, regime-adaptive rule for selecting $\epsilon$ would be a useful direction for future work.

Several substantive extensions remain open. The most natural direction is to generalize the framework to multiple background datasets, where the principal-angle machinery would need to be replaced by a multi-subspace analog along the lines of \citet{feng2018angle}, and where new questions arise about how to define and estimate a coherent shared structure when different background datasets share different aspects of the target. A second extension concerns optimal shrinkage of sample eigenvalues: our procedure improves estimation through better eigenvector estimation, but estimation error could be further reduced by combining this with the shrinkage approaches of \citet{donoho2018optimal} or with task-specific shrinkage as in \citet{sifaou2020high}.

\section{Disclosure statement}\label{disclosure-statement}

The authors have no conflicts of interest to declare. Generative AI (Gemini 3.5 Flash) is used for editing and literature review purposes only. The terms of use of the AI tool have been checked by the authors. Core methodology developments and experiments are conducted by the authors. The authors confirmed the originality and accuracy of the content.

\section{Data Availability Statement}

The stock return data that support the findings of this study are available from Wharton Research Data Services. Restrictions apply to the availability of these data, which were used under license for this study. Data are available at https://wrds-www.wharton.upenn.edu/ with the permission of Wharton Research Data Services. The LGG data that support the findings of this study are openly available in UCSC Xena platform at https://xenabrowser.net/datapages/.

\section{Funding}

Yoon and Ahn's work was partially supported by the National Research Foundation of Korea(NRF) grants (RS-2022-NR068758, RS-2026-25517529). Kim's work was supported by the NRF grant funded by the Korea government(MSIT) (RS-2026-25472832). Jung's work was partially supported by the NRF grant (RS-2024-00333399).

\phantomsection\label{supplementary-material}
\bigskip

\begin{center}

{\large\bf SUPPLEMENTARY MATERIAL}

\end{center}

\begin{description}
\item[Appendix:]
Contains proofs for the theorems and additional simulation study results. (.pdf)
\item[R code:] Contains R codes to implement the proposed method and generate data used in the simulation study. (.zip) 
\end{description}

\bibliography{references}

\makeatletter\@input{xx_Appendix.tex}\makeatother

\end{document}


\maketitle

\section{Algorithm for shared subspace rank and index estimation}\label{sec:appendix-algorithm}

\begin{algorithm}
\caption{Shared subspace rank and index estimation}\label{alg:shared rank}
\begin{algorithmic}[1]
\Require sample eigenvectors $\hQ, \hU$; debiased spikes
  $\{\tlamb_i\}_{i=1}^{r_X}, \{\tgamm_i\}_{i=1}^{r_Y}$; aspect ratios
  $c_X, c_Y$; maximum iterations $\mathrm{maxiter}$
\Ensure shared rank $\hrS$ and index sets $\hPsiX, \hPsiY$
\Statex
\State \textbf{Initialize:} $\hPsiX \gets \{r_X\}$, \enspace
  $\hPsiY \gets \{r_Y\}$, \enspace $\text{iter} \gets 0$
\Repeat
  \State $\hPsiX^{\text{prev}} \gets \hPsiX$, \enspace
    $\hPsiY^{\text{prev}} \gets \hPsiY$
  \State $\text{iter} \gets \text{iter} + 1$
  \State \Comment{Step 1: update the shared rank}
  \State $\text{cutoff} \gets
    \phi_{c_X}\big(\tlamb_{\max(\hPsiX)}\big)\,
    \phi_{c_Y}\big(\tgamm_{\max(\hPsiY)}\big)$
  \State $\hrS \gets \sum_k\mathbb{I}\big(\sigma_k(\hQ^T\hU) \ge \text{cutoff}\big)$
  \State \Comment{Step 2: update the shared index sets}
  \State $\bar P_{\hrS} \gets \Pi_{\hrS}(\hP_X + \hP_Y)$
  \State $\hPsiX \gets \argmax_{i;\,\hrS}
    \big\|\bar P_{\hrS}\,\hP_{X,i}\big\|_F$
  \State $\hPsiY \gets \argmax_{j;\,\hrS}
    \big\|\bar P_{\hrS}\,\hP_{Y,j}\big\|_F$
\Until{$(\hPsiX = \hPsiX^{\text{prev}} \text{ and }
  \hPsiY = \hPsiY^{\text{prev}})$ or $\text{iter} \ge \mathrm{maxiter}$}
\State \Return $\hrS, \hPsiX, \hPsiY$
\end{algorithmic}
\end{algorithm}

\section{Proofs}

\subsection{Useful lemmas}

We collect several technical lemmas used in the proofs of the main results. The first three are standard concentration facts for high-dimensional unit vectors; we include proofs for self-containment. The remaining lemmas are specific to our setting.

\begin{lem}\label{lem:high-dimensional uniform}
Let $z_p$ be a random vector uniformly distributed on $\mathbb{S}^{p-1}$, and let $\{w_p\}_{p=1}^\infty$ be a sequence with $w_p \in \mathbb{S}^{p-1}$ for each $p$. Then
\[
\inner{z_p}{w_p} \aslim 0 \quad \text{as} \quad p \rightarrow \infty.
\]
\end{lem}
\begin{proof}
By Theorem 3.4.5 of \citet{vershynin_2026}, for any $\epsilon > 0$,
\[
\mathbb{P}(|\inner{z_p}{w_p}| > \epsilon) \leq 4 \exp\Big(-\frac{\epsilon^2 p}{2}\Big).
\]
Summing over $p$, $\sum_{p=1}^\infty \mathbb{P}(|\inner{z_p}{w_p}| > \epsilon) < \infty$, and the conclusion follows from the Borel-Cantelli lemma.
\end{proof}

We now show that the same conclusion holds when $z_p$ is constrained to a lower-dimensional subspace.
\begin{lem}\label{lem:random vs fixed}
Let $\calZ$ be a $(p - d_Z)$-dimensional subspace of $\mathbb{R}^p$, with $d_Z$ a fixed constant, and let $z_p$ be a random vector uniformly distributed on the unit sphere in $\calZ$, i.e. $\mathbb{S}^{p-1} \cap \calZ$. Then, for any sequence $\{w_p\}_{p=1}^\infty$ with $w_p \in \mathbb{S}^{p-1}$ for each $p$,
\[
\inner{z_p}{w_p} \aslim 0 \quad \text{as} \quad p \to \infty.
\]
\end{lem}
\begin{proof}
Let $P_Z$ be the orthogonal projection onto $\calZ$. Decomposing $w_p = P_Z w_p + P_Z^{\perp} w_p$ and using $z_p \in \calZ$,
\[
\inner{z_p}{w_p} = \inner{z_p}{P_Z w_p} = \|P_Z w_p\| \, \inner{z_p}{w_{p,Z}},
\]
where $w_{p,Z} = P_Z w_p / \|P_Z w_p\|$. Since $w_{p,Z}$ is a unit vector in $\calZ$ and $\|P_Z w_p\| \leq 1$, Lemma~\ref{lem:high-dimensional uniform} applied within $\calZ$ gives $\inner{z_p}{w_{p,Z}} \aslim 0$, which yields the conclusion.
\end{proof}

The result of Lemma~\ref{lem:random vs fixed} generalizes to the case where $w_p$ is also uniformly random.
\begin{lem}\label{lem:random vs random}
Let $\calZ$ and $\calW$ be subspaces of $\mathbb{R}^p$ with dimensions $p - d_Z$ and $p - d_W$ respectively, where $d_Z$ and $d_W$ are fixed constants. Let $z_p$ and $w_p$ be independent random vectors, uniformly distributed on the unit spheres in $\calZ$ and $\calW$ respectively, i.e.  $\mathbb{S}^{p-1} \cap \calZ$ and  $\mathbb{S}^{p-1} \cap \calW$. Then
\[
\inner{z_p}{w_p} \aslim 0 \quad \text{as} \quad p \to \infty.
\]
\end{lem}
\begin{proof}
For a fixed $p$ larger than $\max(d_Z,d_W)$, we have
\[
\begin{aligned}
\mathbb{P}(|\inner{z_p}{w_p}| > \epsilon|w_p) &= \mathbb{P}(\|P_Zw_p\||\inner{z_p}{w_p/\|P_Zw_p\|}| > \epsilon|w_p) \\ & \leq \mathbb{P}(|\inner{z_p}{w_p/\|P_Zw_p\|}| > \epsilon|w_p) \\ &\leq 4\exp(-\epsilon^2(p-d_Z)/2).
\end{aligned}
\]
Here, the first inequality holds since $\|P_Zw_p\| \leq 1$ and we are using the technique from the proof of Lemma~\ref{lem:random vs fixed}. Then, by taking the expectation with respect to $w_p$, law of total expectation gives
\[
\mathbb{P}(|\inner{z_p}{w_p}| > \epsilon) \leq 4\exp(-\epsilon^2(p-d_Z)/2),
\]
and the conclusion follows from the Borel-Cantelli lemma.
\end{proof}
Lemma~\ref{lem:random vs random} confirms the intuition that two independent random vectors in high dimensions are nearly orthogonal. The next lemma describes the eigenstructure of the sum of two orthogonal projection matrices.

\begin{lem}\label{lem:projection sum}
Let $\calZ$ and $\calW$ be subspaces of $\Real^p$ with dimensions $d_Z$ and $d_W$, and let $P_Z = Z Z^\top$ and $P_W = W W^\top$, where $Z$ and $W$ are orthonormal bases of $\calZ$ and $\calW$ respectively. Let $\theta_1, \ldots, \theta_{\min(d_Z, d_W)}$ denote the principal angles between $\calZ$ and $\calW$, and let $z_1, \ldots, z_{d_Z}$ and $w_1, \ldots, w_{d_W}$ be principal vectors forming orthonormal bases of $\calZ$ and $\calW$ such that $z_i^\top w_j = 0$ for $i \ne j$ and $z_k^\top w_k = \cos\theta_k$ for $k = 1, \ldots, \min(d_Z, d_W)$. Then the nonzero eigenvalues of $P_Z + P_W$ are
\[
\ell(P_Z + P_W) = 
\begin{cases}
\{1 \pm \cos\theta_k\}_{k=1}^{d_Z}, & d_Z = d_W, \\
\{1 \pm \cos\theta_k\}_{k=1}^{\min(d_Z, d_W)} \cup \{\underbrace{1, \ldots, 1}_{|d_Z - d_W|}\}, & d_Z \ne d_W,
\end{cases}
\]
and the corresponding (unnormalized) eigenvectors are $z_k \pm w_k$ for $k = 1, \ldots, \min(d_Z, d_W)$, together with the additional principal vectors of the larger subspace when $d_Z \ne d_W$.
\end{lem}

\begin{proof}
Since $\{z_k\}$ and $\{w_k\}$ form orthonormal bases of $\calZ$ and $\calW$,
$
P_Z = \sum_{k=1}^{d_Z} z_k z_k^\top$ and $P_W = \sum_{k=1}^{d_W} w_k w_k^\top.
$
Consider first the case $d_Z = d_W = d$. The principal-vector identities $z_i^\top w_j = 0$ for $i \ne j$ and $z_k^\top w_k = \cos\theta_k$ give
\[
P_Z w_k = \cos\theta_k \, z_k, \qquad P_W z_k = \cos\theta_k \, w_k, \qquad P_Z z_k = z_k, \qquad P_W w_k = w_k.
\]
Therefore, for $k = 1, \ldots, d$, $
(P_Z + P_W)(z_k \pm w_k) = (1 \pm \cos\theta_k)(z_k \pm w_k).
$
When $d_Z \ne d_W$, suppose without loss of generality that $d_Z < d_W$. The principal vectors $w_{d_Z+1}, \ldots, w_{d_W}$ lie in $\calW$ and are orthogonal to every $z_k$ (corresponding to principal angles of $\pi/2$), so $P_Z w_l = 0$ for $l = d_Z + 1, \ldots, d_W$. The previous computation gives $(P_Z + P_W)(z_k \pm w_k) = (1 \pm \cos\theta_k)(z_k \pm w_k)$ for $k = 1, \ldots, d_Z$, while
\[
(P_Z + P_W) w_l = P_W w_l = w_l \quad \text{for } l = d_Z + 1, \ldots, d_W.
\]
\end{proof}

The last lemma is used in the proofs of Theorems~\ref{thm:index consistency} and \ref{thm:Frobenius performance}. It states that almost-sure convergence in Frobenius distance transfers cleanly through differences and products of orthogonal projections.
\begin{lem}\label{lem:matrix function asymptotics}
For each $p \geq 1$, let $\calP_r^p$ denote the set of $p \times p$ orthogonal projection matrices of rank $r$. Let $A_n, B_n \in \calP_{r_1}^{p(n)}$ and $C_n \in \calP_{r_2}^{p(n)}$ be sequences of random projection matrices, where $p(n) \to \infty$ as $n \to \infty$. Suppose $\|A_n - B_n\|_F \aslim 0$. Then:
\begin{enumerate}[label=(\roman*)]
    \item if $\|A_n - C_n\|_F \aslim \kappa_1$ for a constant $\kappa_1$, then $\|B_n - C_n\|_F \aslim \kappa_1$;
    \item if $\|A_n C_n\|_F \aslim \kappa_2$ for a constant $\kappa_2$, then $\|B_n C_n\|_F \aslim \kappa_2$.
\end{enumerate}
\end{lem}

\begin{proof}
We prove (i); the proof of (ii) is analogous. The Frobenius norm is $1$-Lipschitz in each argument, so
\[
\big| \|A_n - C_n\|_F - \|B_n - C_n\|_F \big| \leq \|A_n - B_n\|_F.
\]
Since $\|A_n - B_n\|_F \aslim 0$, we have $\big| \|A_n - C_n\|_F - \|B_n - C_n\|_F \big| \aslim 0$. Combined with $\|A_n - C_n\|_F \aslim \kappa_1$, this gives $\|B_n - C_n\|_F \aslim \kappa_1$.

For (ii), the analogous inequality is $\big| \|A_n C_n\|_F - \|B_n C_n\|_F \big| \leq \|(A_n - B_n) C_n\|_F \leq \|A_n - B_n\|_F \|C_n\|_{\mathrm{op}}$, where $\|C_n\|_{\mathrm{op}} = 1$ since $C_n$ is an orthogonal projection.
\end{proof}

\subsection{Proof of Lemma \ref{lem:sample inner product}}

\begin{proof}
Decompose each sample eigenvector into its signal and noise components:
\[
\hat v_{i,1} = P_1 \hat v_{i,1} + P_1^\perp \hat v_{i,1}, \qquad \hat v_{j,2} = P_2 \hat v_{j,2} + P_2^\perp \hat v_{j,2},
\]
where $P_k = \sum_{i=1}^{r_k} v_{i,k} v_{i,k}^\top$ is the projection onto the span of the true spiked eigenvectors of dataset $k$ and $P_k^\perp = I_p - P_k$. The inner product then decomposes as
\[
\inner{\hat v_{i,1}}{\hat v_{j,2}} = \inner{P_1 \hat v_{i,1}}{P_2 \hat v_{j,2}} + \inner{P_1^\perp \hat v_{i,1}}{P_2 \hat v_{j,2}} + \inner{P_1 \hat v_{i,1}}{P_2^\perp \hat v_{j,2}} + \inner{P_1^\perp \hat v_{i,1}}{P_2^\perp \hat v_{j,2}}.
\]
We analyze each term in turn. For the signal-signal term,
\[
\inner{P_1 \hat v_{i,1}}{P_2 \hat v_{j,2}} = \sum_{l=1}^{r_1} \sum_{m=1}^{r_2} \inner{v_{l,1}}{\hat v_{i,1}} \inner{v_{m,2}}{\hat v_{j,2}} \inner{v_{l,1}}{v_{m,2}}.
\]
By Proposition~\ref{prop:classic asymptotics}, $|\inner{v_{l,1}}{\hat v_{i,1}}| \aslim \phi_{c_1}(\ell_{i,1}) \delta_{l,i}$ and $|\inner{v_{m,2}}{\hat v_{j,2}}| \aslim \phi_{c_2}(\ell_{j,2}) \delta_{m,j}$ as $n_1, n_2 \to \infty$, so only the $(l, m) = (i, j)$ term survives:
\[
\big| \inner{P_1 \hat v_{i,1}}{P_2 \hat v_{j,2}} \big| \aslim \phi_{c_1}(\ell_{i,1}) \phi_{c_2}(\ell_{j,2}) \, |\inner{v_{i,1}}{v_{j,2}}|.
\]

For the signal-noise terms, by Theorem~6 of \citet{paul2007asymptotics}, $P_1^\perp \hat v_{i,1} / \|P_1^\perp \hat v_{i,1}\|$ is uniformly distributed on the unit sphere in $\mathrm{range}(P_1^\perp)$ and is independent of $\|P_1^\perp \hat v_{i,1}\|$; an analogous statement holds for $P_2^\perp \hat v_{j,2}$. Expanding
\[
\inner{P_1^\perp \hat v_{i,1}}{P_2 \hat v_{j,2}} = \sum_{m=1}^{r_2} \inner{v_{m,2}}{\hat v_{j,2}} \inner{v_{m,2}}{P_1^\perp \hat v_{i,1}},
\]
the inner products $\inner{v_{m,2}}{\hat v_{j,2}}$ converge by Proposition~\ref{prop:classic asymptotics}, and the inner products $\inner{v_{m,2}}{P_1^\perp \hat v_{i,1}}$ converge to zero by Lemma~\ref{lem:random vs fixed}. Hence $\inner{P_1^\perp \hat v_{i,1}}{P_2 \hat v_{j,2}} \aslim 0$. The same argument applied to the symmetric counterpart gives $\inner{P_1 \hat v_{i,1}}{P_2^\perp \hat v_{j,2}} \aslim 0$.

For the noise-noise term, $\inner{P_1^\perp \hat v_{i,1}}{P_2^\perp \hat v_{j,2}} \aslim 0$ by Lemma~\ref{lem:random vs random}.
\end{proof}

\subsection{Proof of Theorem~\ref{thm:singular convergence}}
\begin{proof}
Since singular values are invariant under column permutation,
\[
\sigma(\hQ^\top \hU) = \sigma\big([\hQ_S, \hQ_D]^\top [\hU_S, \hU_D]\big) = \sigma\begin{bmatrix} \hQ_S^\top \hU_S & \hQ_S^\top \hU_D \\ \hQ_D^\top \hU_S & \hQ_D^\top \hU_D \end{bmatrix}.
\]
Lemma~\ref{lem:sample inner product} applied to the off-diagonal blocks gives
\begin{equation}\label{eq:shared distinct orthogonal}
\hQ_S^\top \hU_D \aslim 0_{r_S, d_Y}, \quad \hQ_D^\top \hU_S \aslim 0_{d_X, r_S} \quad \text{as } n_X, n_Y \to \infty,
\end{equation}
since $Q_S \perp U_D$ and $Q_D \perp U_S$. For the diagonal blocks, the singular values are unaffected by sign flips of the sample eigenvectors, so we may assume $\mathrm{sign}(\inner{\hat q_i}{\hat u_j}) = \mathrm{sign}(\inner{q_i}{u_j})$. Lemma~\ref{lem:sample inner product} then gives
\begin{equation}\label{eq:shared convergence}
\hQ_S^\top \hU_S \aslim \Phi_{X,S} Q_S^\top U_S \Phi_{Y,S} = \Phi_{X,S} R \Phi_{Y,S} \quad \text{as } n_X, n_Y \to \infty,
\end{equation}
and
\begin{equation}\label{eq:distinct convergence}
\hQ_D^\top \hU_D \aslim \Phi_{X,D} Q_D^\top U_D \Phi_{Y,D} \quad \text{as } n_X, n_Y \to \infty.
\end{equation}
Define the block-diagonal limit
\[
G = \begin{bmatrix} \Phi_{X,S} R \Phi_{Y,S} & 0_{r_S, d_Y} \\ 0_{d_X, r_S} & \Phi_{X,D} Q_D^\top U_D \Phi_{Y,D} \end{bmatrix}.
\]
By continuity of singular values, $\sigma(\hQ^\top \hU) \aslim \sigma(G)$ as $n_X, n_Y \to \infty$. Since $G$ is block diagonal, $\sigma(G)$ is the union of the singular values of its two diagonal blocks, which gives the stated result.
\end{proof}

\subsection{Proof of Lemma \ref{lem:rotation inequality}}

\begin{proof}
    For $A,B \in \Real^{n\times n}$, we have 
    \[
    \sigma_i(AB) \geq \sigma_n(A)\sigma_i(B), \ i=1,\ldots,n.
    \]
    Applying this property, we can directly obtain that
    \[
    \begin{aligned}
    \sigma_{r_S}(\Phi_{X,S}R\Phi_{Y,S}) &\geq \sigma_{r_S}(\Phi_{X,S}R) \sigma_{r_S}(\Phi_{Y,S}) \\ &=\sigma_{r_S}(\Phi_{X,S})\sigma_{r_S} (\Phi_{Y,S}) \\ &= \phi_{c_X}(\lambda_{r_S,S}) \phi_{c_Y}(\gamma_{r_S,S}).
    \end{aligned}
    \]
\end{proof}

\subsection{Proof of Theorem~\ref{thm:rank consistency}}
\begin{proof}
Under Assumption~\ref{ass:above transition}, the bias correction in Proposition~\ref{prop:classic asymptotics} yields $\tilde\lambda_{r_S, S} \aslim \lambda_{r_S, S}$ and $\tilde\gamma_{r_S, S} \aslim \gamma_{r_S, S}$ as $n_X, n_Y \to \infty$. By Theorem~\ref{thm:singular convergence}, the smallest shared singular value $\sigma_{r_S}(\hQ_S^\top \hU_S)$ converges almost surely to $\sigma_{r_S}(\Phi_{X,S} R \Phi_{Y,S})$, and Lemma~\ref{lem:rotation inequality} bounds this from below by $\phi_{c_X}(\lambda_{r_S, S}) \phi_{c_Y}(\gamma_{r_S, S})$. Assumption~\ref{ass:shared distinct separable} ensures the cutoff $\phi_{ c_X}(\tilde\lambda_{r_S, S}) \phi_{c_Y}(\tilde\gamma_{r_S, S})$ separates the shared singular values from the distinct ones, giving $\hrS^* \aslim r_S$.
\end{proof}

\subsection{Proof of Theorem~\ref{thm:index consistency}}
\begin{proof}
Define
\[
\hM_0 = \hQ_S \hQ_S^\top + \hU_S \hU_S^\top + \hQ_D \hQ_D^\top + \hU_D \hU_D^\top = \hM_S + \hM_D,
\]
where $\hM_S = \hQ_S \hQ_S^\top + \hU_S \hU_S^\top$ and $\hM_D = \hQ_D \hQ_D^\top + \hU_D \hU_D^\top$. The proof proceeds in two steps: (i) the top-$r_S$ eigenprojection of $\hM_0$ converges to that of $\hM_S$, and (ii) on the limit, projection onto the eigenspace separates indices in $\Psi_X$ from indices outside it.

\textbf{Step 1: asymptotic equality.}
We first establish
\begin{equation}\label{eq:total equal partial}
\|\bar P_{r_S} - \Pi_{r_S}(\hM_S)\|_F \aslim 0 \quad \text{as } n_X, n_Y \to \infty.
\end{equation}
Let $\hV_S = \calV_{r_S}(\hM_S)$ where $\calV_{r_S}$ returns the matrix whose columns are top $r_S$ eigenvectors and write $\hQ_S^\top \hU_S = \hA \hD \hB^\top$ via singular value decomposition. By Lemma~\ref{lem:projection sum}, the $k$-th column of $\hQ_S \hA + \hU_S \hB$ is an unnormalized eigenvector of $\hM_S$ corresponding to eigenvalue $1 + \sigma_k(\hQ_S^\top \hU_S)$ for $k = 1, \ldots, r_S$, so
\begin{equation}\label{eq:shared eigenvector}
\hV_S = (\hQ_S \hA + \hU_S \hB) \hN^{-1}, \qquad \hN = \big( 2(\hD + I_{r_S}) \big)^{1/2}.
\end{equation}
Theorem~3.4 of Chapter~5 in \citet{stewart1990matrix} gives
\[
\|\bar P_{r_S} - \Pi_{r_S}(\hM_S)\|_F \leq \frac{\sqrt{2} \|\hR\|_F}{\hat\delta},
\]
where $\hR = \hM_0 \hV_S - \hV_S \hV_S^\top \hM_S \hV_S = \hM_D \hV_S$ and $\hat\delta = \min |(\ell(\hM_0) \setminus \ell_{1:r_S}(\hM_0)) - \ell_{1:r_S}(\hM_S)|$ is the eigengap, with $\ell_{1:r_S}$ denoting the set of top $r_S$ eigenvalues. That is, $\hat{\delta}$ implies the minimal absolute difference between elements of two eigenvalue sets. Equations~\eqref{eq:shared distinct orthogonal} and \eqref{eq:shared eigenvector} give $\|\hR\|_F \aslim 0$. By \eqref{eq:shared distinct orthogonal} and Lemma~\ref{lem:projection sum}, assuming $d_X = d_Y$ without loss of generality,
\[
\ell(\hM_0) \aslim \{1 \pm \sigma_k(\Phi_{X,S} R \Phi_{Y,S})\}_{k=1}^{r_S} \cup \{1 \pm \sigma_k(\Phi_{X,D} Q_D^\top U_D \Phi_{Y,D})\}_{k=1}^{d_X},
\]
and Assumption~\ref{ass:shared distinct separable} yields
\[
\hat\delta \aslim \sigma_{r_S}(\Phi_{X,S} R \Phi_{Y,S}) - \sigma_1(\Phi_{X,D} Q_D^\top U_D \Phi_{Y,D}) > 0,
\]
which combined with $\|\hR\|_F \aslim 0$ proves the claim.

\textbf{Step 2: index separation.}
Since sign flips of the sample eigenvectors do not affect $\hP_{X,i}$, we may assume $\mathrm{sign}(\inner{\hat q_i}{\hat u_j}) = \mathrm{sign}(\inner{q_i}{u_j})$. From $\Pi_{r_S}(\hM_S) = \hV_S \hV_S^\top$ and \eqref{eq:shared eigenvector},
\[
\|\Pi_{r_S}(\hM_S) \hP_{X,i}\|_F = \sqrt{\hat q_i^\top (\hQ_S \hA + \hU_S \hB) \hN^{-2} (\hQ_S \hA + \hU_S \hB)^\top \hat q_i}.
\]
If $i \notin \Psi_X$, Lemma~\ref{lem:sample inner product} gives $(\hQ_S \hA + \hU_S \hB)^\top \hat q_i \aslim 0_{r_S}$, hence $\|\Pi_{r_S}(\hM_S) \hP_{X,i}\|_F \aslim 0$.

If $i \in \Psi_X$, Lemma~\ref{lem:sample inner product} gives
\[
\hN^{-1} (\hQ_S \hA + \hU_S \hB)^\top \hat q_i \aslim N^{-1} (A^\top e_i + B^\top \rho_i),
\]
where $N = (2(D + I_{r_S}))^{1/2}$ with $\Phi_{X,S} R \Phi_{Y,S} = A D B^\top$; $e_i$ is the $r_S$-dimensional vector with a single $1$ at the coordinate where $\hat q_{l, r_S} = \hat q_i$; and $\rho_i$ has $k$-th entry $\phi_{c_X}(\lambda_i) \phi_{c_Y}(\gamma_{k, S}) \inner{q_i}{u_{k, S}}$. Therefore $\|\Pi_{r_S}(\hM_S) \hP_{X,i}\|_F \aslim \mu_i > 0$, where
\[
\mu_i = \sqrt{(A^\top e_i + B^\top \rho_i)^\top N^{-2} (A^\top e_i + B^\top \rho_i)}.
\]
Combining the two cases,
\[
\|\Pi_{r_S}(\hM_S) \hP_{X,i}\|_F \aslim \begin{cases} \mu_i & i \in \Psi_X, \\ 0 & i \notin \Psi_X, \end{cases}
\]
and \eqref{eq:total equal partial} together with Lemma~\ref{lem:matrix function asymptotics} extends this to $\|\bar P_{r_S} \hP_{X,i}\|_F$, yielding $\hPsiX^* \aslim \Psi_X$. The argument for $\hPsiY^*$ is analogous.
\end{proof}

\subsection{Proof of Theorem \ref{thm:Frobenius performance}}\label{sec:appendix-proof}

\begin{proof}
Let $\hM_0^{(\alpha)} = \alpha \hP_X + (1 - \alpha) \hP_Y$ and $\hM_S^{(\alpha)} = \alpha \hP_{X,S} + (1 - \alpha) \hP_{Y,S}$. The argument from the proof of Theorem~\ref{thm:index consistency} extends with minor modifications to give
\begin{equation}\label{eq:total equal parital weighted}
\|\hP_S^{(\alpha)} - \Pi_{r_S}(\hM_S^{(\alpha)})\|_F \aslim 0 \quad \text{as } n_X, n_Y \to \infty.
\end{equation}
The only substantive change is that Lemma~\ref{lem:projection sum} is replaced by its weighted analog: in the notation of that lemma,
\begin{equation}\label{eq:projection sum general}
\ell(\alpha P_Z + (1 - \alpha) P_W) =
\begin{cases}
\big\{ \tfrac{1 \pm g(\cos\theta_k, \alpha)}{2} \big\}_{k=1}^{d_Z}, & d_Z = d_W, \\
\big\{ \tfrac{1 \pm g(\cos\theta_k, \alpha)}{2} \big\}_{k=1}^{d_W} \cup \{\underbrace{\alpha, \ldots, \alpha}_{d_Z - d_W}\}, & d_Z > d_W, \\
\big\{ \tfrac{1 \pm g(\cos\theta_k, \alpha)}{2} \big\}_{k=1}^{d_Z} \cup \{\underbrace{1 - \alpha, \ldots, 1 - \alpha}_{d_W - d_Z}\}, & d_Z < d_W,
\end{cases}
\end{equation}
where $g(c, \alpha) = \sqrt{1 - 4\alpha(1 - \alpha)(1 - c^2)}$, with eigenvectors $(\alpha \cos\theta_k) z_k + (\lambda_k^\pm - \alpha) w_k$ corresponding to the eigenvalues $\lambda_k^\pm = (1 \pm g(\cos\theta_k, \alpha))/2$, for $k = 1, \ldots, \min(d_Z, d_W)$. This generalizes Lemma~\ref{lem:projection sum} and follows from the cosine-sine (CS) decomposition \citep[Chapter~1, Section~5.1]{stewart1990matrix}.

Combining \eqref{eq:total equal parital weighted} with Lemma~\ref{lem:matrix function asymptotics}, it suffices to characterize the limit of $\tfrac{1}{2}\|P_S - \Pi_{r_S}(\hM_S^{(\alpha)})\|_F^2$. Let $\hV_S^{(\alpha)} = \calV_{r_S}(\hM_S^{(\alpha)})$ and write $\hQ_S^\top \hU_S = \hA \hD \hB^\top$ via singular value decomposition. Then
\[
\tfrac{1}{2}\|P_S - \Pi_{r_S}(\hM_S^{(\alpha)})\|_F^2 = r_S - \tr\big(Q_S^\top (\hQ_S \hA \hW_X + \hU_S \hB \hW_Y) \hN_W^{-2} (\hQ_S \hA \hW_X + \hU_S \hB \hW_Y)^\top Q_S\big),
\]
where $\hW_X = \alpha \hD$, $\hW_Y = \tfrac{1}{2}(I_{r_S} + g(\hD, \alpha)) - \alpha I_{r_S}$, and $\hN_W = (\hW_X^\top \hW_X + \hW_Y^\top \hW_Y + \hW_X^\top \hD \hW_Y + \hW_Y \hD \hW_X)^{1/2}$, with $g(\hD, \alpha) = \tdiag(g(\hat d_1, \alpha), \ldots, g(\hat d_{r_S}, \alpha))$ and $\hat d_k = \sigma_k(\hQ_S^\top \hU_S)$. Writing $\Phi_{X,S} R \Phi_{Y,S} = A D B^\top$,
\begin{equation}\label{eq:exact performance}
L(\alpha) = r_S - \tr\big((\Phi_{X,S} A W_X + R \Phi_{Y,S} B W_Y) N_W^{-2} (W_X A^\top \Phi_{X,S} + W_Y B^\top \Phi_{Y,S} R^\top)\big),
\end{equation}
where $W_X, W_Y, N_W$ are defined as above with $\hD$ replaced by $D$.

The exact form \eqref{eq:exact performance} is opaque, but the top-$r_S$ eigenvalues of $\hM_S^{(\alpha)}$ dominate the rest when $\sigma_k(\hQ_S^\top \hU_S)$ is close to $1$ (visible from \eqref{eq:projection sum general}), so $\Pi_{r_S}(\hM_S^{(\alpha)}) \approx \hM_S^{(\alpha)}$. We therefore approximate $L(\alpha)$ by the limit of $\tfrac{1}{2}\|P_S - \hM_S^{(\alpha)}\|_F^2$. Writing $\beta = 1 - \alpha$ and $\inner{A}{B}_F = \tr(A^\top B)$,
\[
\tfrac{1}{2}\|P_S - \hM_S^{(\alpha)}\|_F^2 = \tfrac{\alpha^2}{2}\|P_S - \hP_{X,S}\|_F^2 + \tfrac{\beta^2}{2}\|P_S - \hP_{Y,S}\|_F^2 + \alpha\beta \inner{P_S - \hP_{X,S}}{P_S - \hP_{Y,S}}_F.
\]
We analyze the three terms.

For the first, $\tfrac{1}{2}\|P_S - \hP_{X,S}\|_F^2 = r_S - \tr(\hQ_S^\top Q_S Q_S^\top \hQ_S)$, and Proposition~\ref{prop:classic asymptotics} gives $\tr(\hQ_S^\top Q_S Q_S^\top \hQ_S) \aslim \sum_{i=1}^{r_S} \phi_{c_X}^2(\lambda_{i,S})$ as $n_X \to \infty$, so
\[
\tfrac{\alpha^2}{2}\|P_S - \hP_{X,S}\|_F^2 \aslim \alpha^2 \Big( r_S - \sum_{i=1}^{r_S} \phi_{c_X}^2(\lambda_{i,S}) \Big).
\]
The same argument applied to the second term gives
\[
\tfrac{\beta^2}{2}\|P_S - \hP_{Y,S}\|_F^2 \aslim \beta^2 \Big( r_S - \sum_{i=1}^{r_S} \phi_{c_Y}^2(\gamma_{i,S}) \Big).
\]
The third term expands as
\[
\inner{P_S - \hP_{X,S}}{P_S - \hP_{Y,S}}_F = r_S - \tr(P_S \hP_{X,S}) - \tr(P_S \hP_{Y,S}) + \tr(\hP_{X,S} \hP_{Y,S}).
\]
The first two trace terms converge as above, and \eqref{eq:shared convergence} gives $\tr(\hP_{X,S} \hP_{Y,S}) = \tr(\hU_S^\top \hQ_S \hQ_S^\top \hU_S) \aslim \tr(\Phi_{X,S}^2 R \Phi_{Y,S}^2 R^\top)$ as $n_X, n_Y \to \infty$. Collecting the convergence results and rewriting them complete the proof.
\end{proof}

\section{Data generation for Section~\ref{subsec: sim beyond pcs}}\label{appendix: data gen}

This appendix details the data-generating processes for Models~I--IV used in the robustness simulations. In all four models, eigenvalues follow the mixed configuration (b) of Section~\ref{subsec:est-sim}.

\textbf{Models II and IV} ($Q_S = U_S$, distinct components absent or present): the shared eigenvectors are generated as in Section~\ref{subsec:est-sim} but without the random rotation, so $U_S = Q_S$.

\textbf{Model III} ($\mathrm{col}(Q_S) = \mathrm{col}(U_S)$, distinct components absent): the shared eigenvectors are generated exactly as in Section~\ref{subsec:est-sim}, with $U_S = Q_S R$ for a random rotation $R$.

\textbf{Model I} (shared subspaces approximately aligned, distinct components present): $Q_S$ is generated as in Section~\ref{subsec:est-sim}. To construct $U_S$ with principal angles $\theta = \pi/36 = 5^\circ$ to $Q_S$, we sample a $p \times r_S$ matrix $Z$ with i.i.d.\ $N(0, 1)$ entries, set $Z_\perp = (I_p - Q_S Q_S^\top) Z$, and let $Q_{S, \perp} \in \calO^{p, r_S}$ be an orthonormal basis of $\mathrm{col}(Z_\perp)$. Then
\[
U_S = Q_S \cos\theta + Q_{S, \perp} \sin\theta.
\]

In Models~I and II, the distinct eigenvectors $Q_D$ and $U_D$ are generated under setting (i) of Section~\ref{subsec:est-sim} (orthogonal distinct subspaces). In Models~III and IV, $d_X = d_Y = 0$, so there are no distinct components.

\section{Additional simulation studies}

\subsection{Simulation study for Theorems~\ref{thm:singular convergence}, \ref{thm:rank consistency}, and~\ref{thm:index consistency}}

\begin{figure}[!ht]
\centering
\begin{subfigure}{.4\textwidth}
  \centering
  \includegraphics[width=1\linewidth]{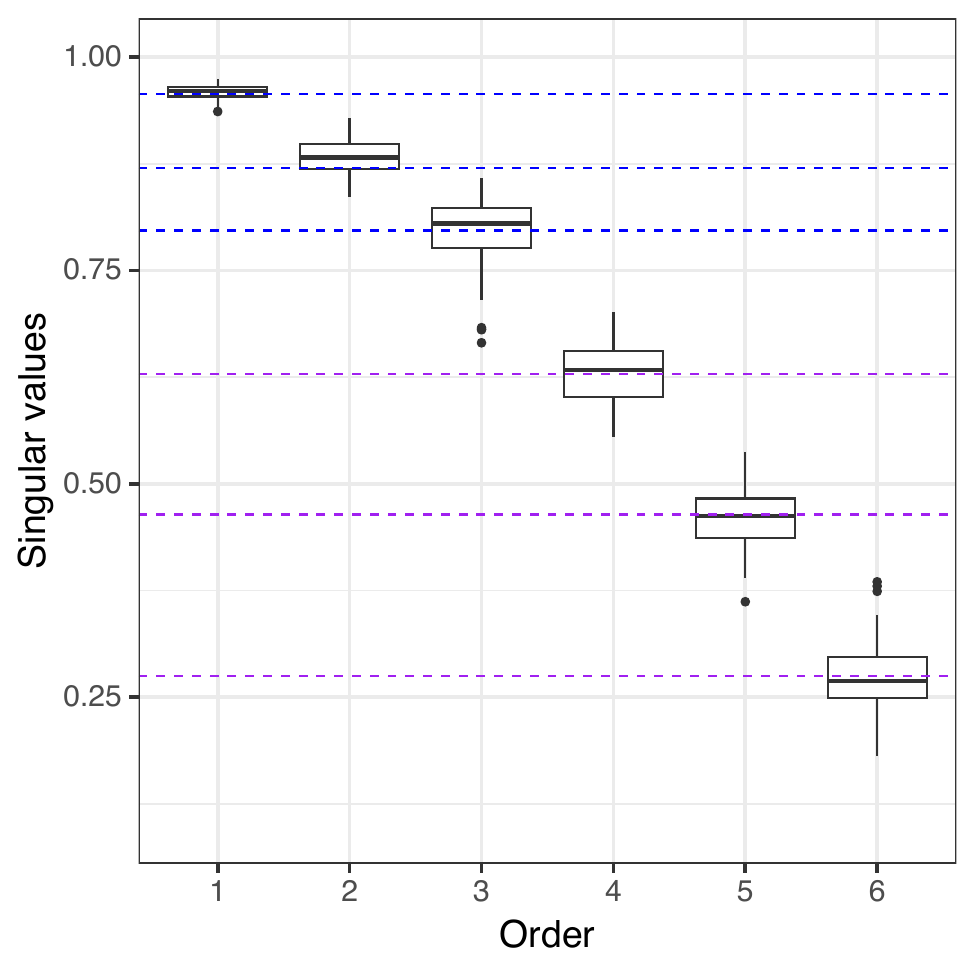}
  \caption{$m = 2$}
\end{subfigure}%
\begin{subfigure}{.4\textwidth}
  \centering
  \includegraphics[width=1\linewidth]{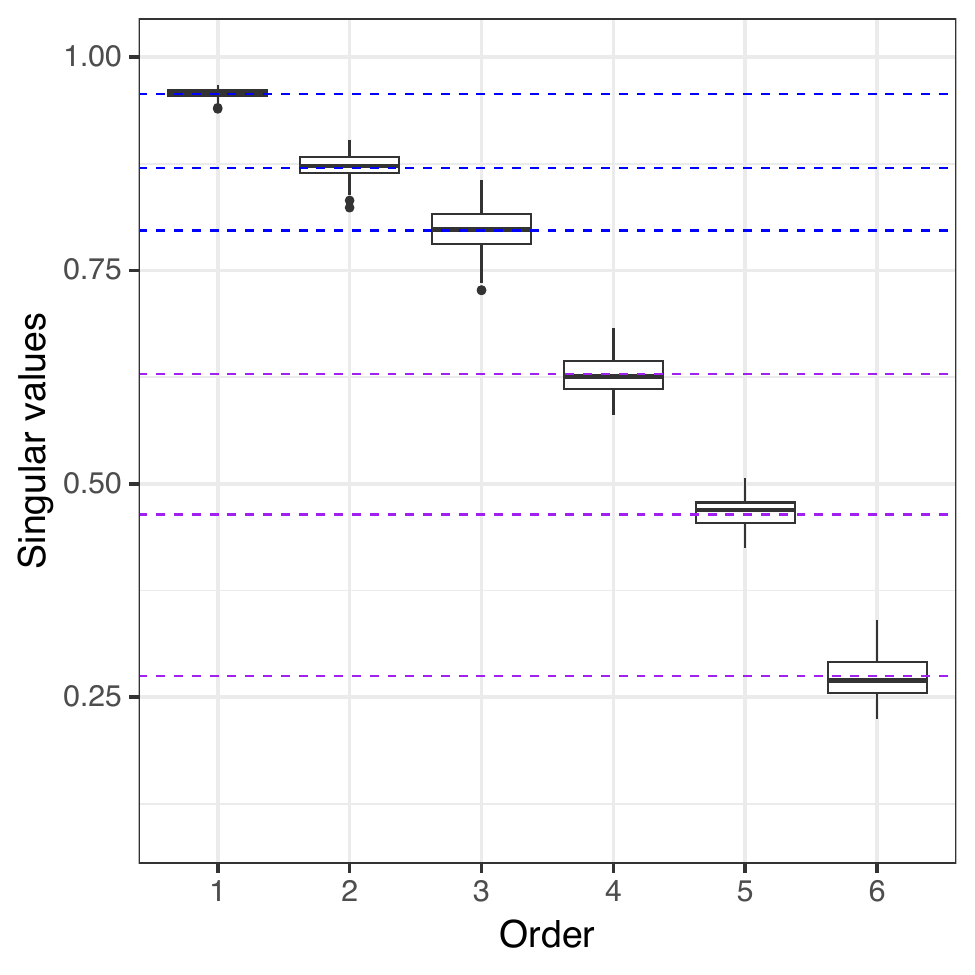}
  \caption{$m = 5$}
\end{subfigure}
\begin{subfigure}{.4\textwidth}
  \centering
  \includegraphics[width=1\linewidth]{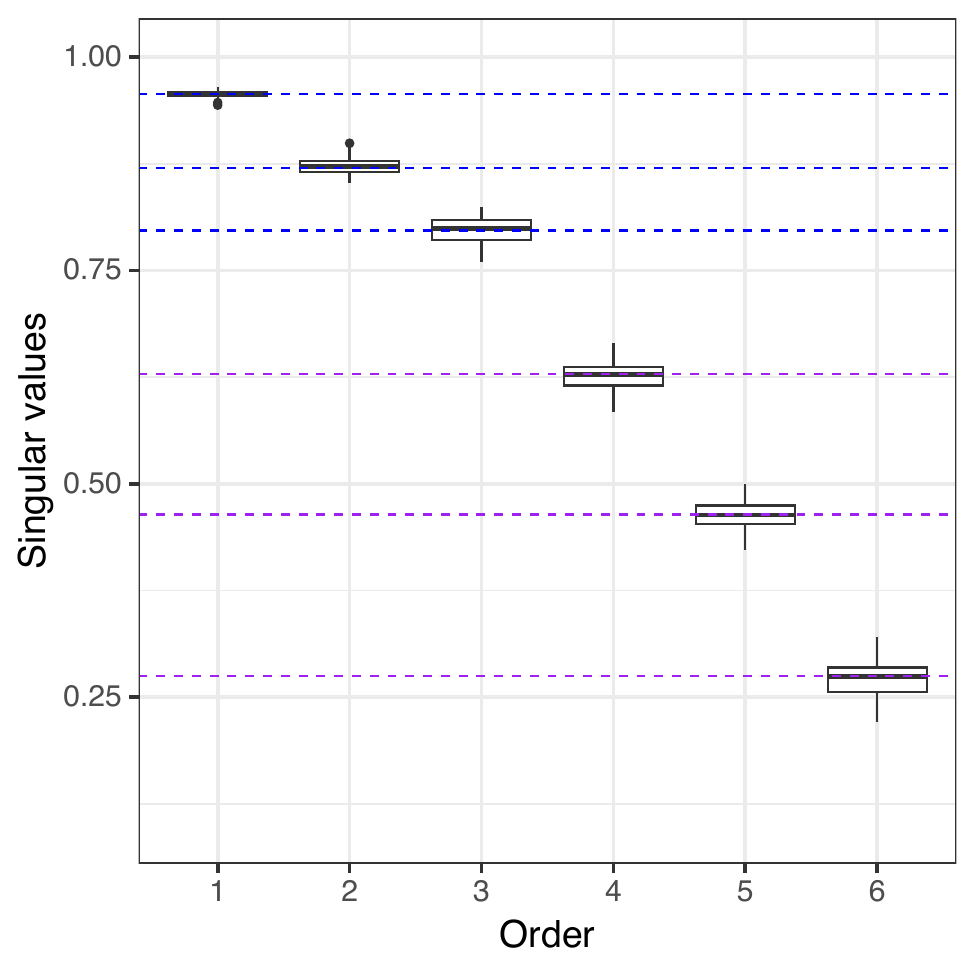}
  \caption{$m = 10$}
\end{subfigure}
\begin{subfigure}{.4\textwidth}
  \centering
  \includegraphics[width=1\linewidth]{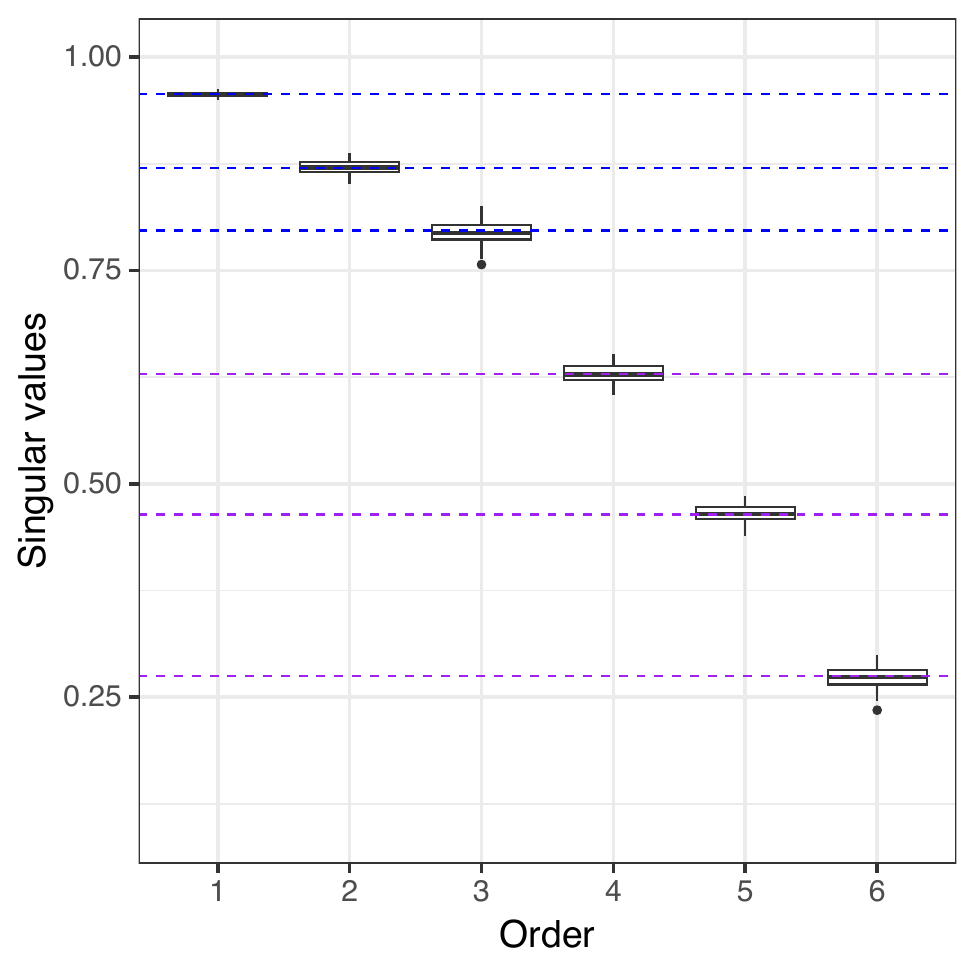}
  \caption{$m = 20$}
\end{subfigure}
\caption{Boxplots of the singular values $\sigma(\hQ^\top \hU)$ over $100$ replications, for $m = 2, 5, 10, 20$. Blue dotted lines mark the asymptotic values $\sigma(\Phi_{X,S} R \Phi_{Y,S})$ (the limits of $\sigma(\hQ_S^\top \hU_S)$); purple dotted lines mark the asymptotic values $\sigma(\Phi_{X,D} Q_D^\top U_D \Phi_{Y,D})$ (the limits of $\sigma(\hQ_D^\top \hU_D)$).}
\label{fig:PA simul}
\end{figure}

We conduct simulations to validate the results of Theorems~\ref{thm:singular convergence}, \ref{thm:rank consistency}, and~\ref{thm:index consistency}. Theorem~\ref{thm:singular convergence} concerns the almost-sure convergence of the principal angles between $\mathrm{span}(\hQ)$ and $\mathrm{span}(\hU)$. To examine this convergence, we set $p = 50m$, $n_X = 30m$, and $n_Y = 60m$, and vary $m$ from $2$ to $20$. For clear visualization, the principal angles between distinct subspaces are kept exceptionally small, with $\sigma(Q_D^\top U_D) = \{0.7, 0.5, 0.3\}$. The shared and distinct dimensions are $r_S = 3$, $d_X = 3$, and $d_Y = 4$, the eigenvalues are $\Lambda = \mathrm{diag}(26, 21, 17, 13, 9, 5)$ and $\Gamma = \mathrm{diag}(39, 34, 30, 23, 19, 11, 4)$, the index sets are $\Psi_X = \{1, 3, 6\}$ and $\Psi_Y = \{1, 4, 7\}$, and $\tau^2 = \eta^2 = 1$.

Figure~\ref{fig:PA simul} shows boxplots of $\sigma(\hQ^\top \hU)$ over $100$ replications for $m = 2, 5, 10, 20$. The blue dotted lines mark the asymptotic values $\sigma(\Phi_{X,S} R \Phi_{Y,S})$ (the limit of $\sigma(\hQ_S^\top \hU_S)$), and the purple dotted lines mark the asymptotic values $\sigma(\Phi_{X,D} Q_D^\top U_D \Phi_{Y,D})$ (the limit of $\sigma(\hQ_D^\top \hU_D)$). The empirical sample singular values are centered around their respective asymptotic values across all values of $m$, including $m = 2$, and the interquartile ranges shrink markedly as $m$ grows, consistent with Theorem~\ref{thm:singular convergence}.

\begin{figure}[!ht]
    \centering
    \begin{subfigure}{.49\textwidth}
  \centering
  \includegraphics[width=1\linewidth]{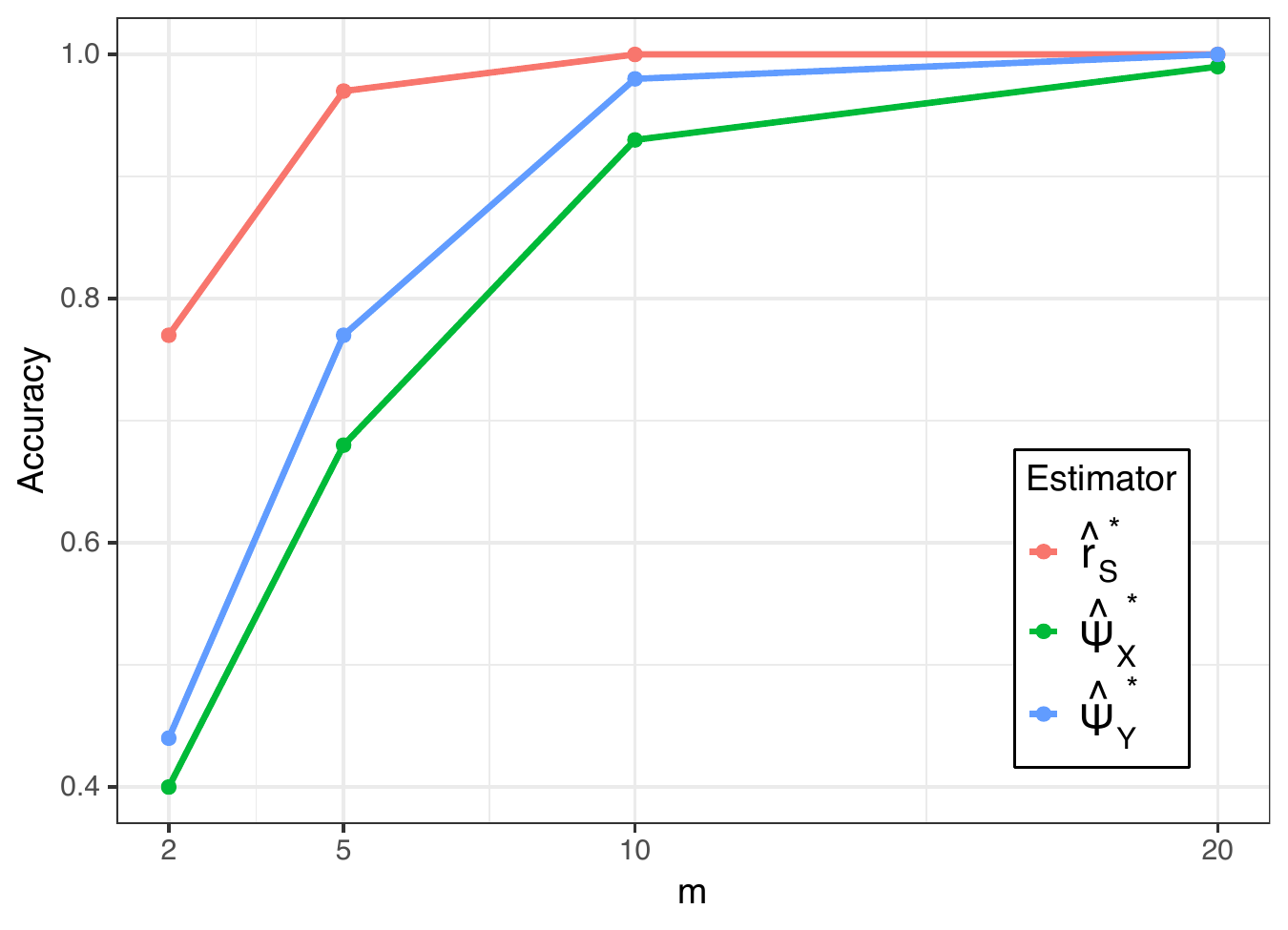}
  \caption{Accuracy}
\end{subfigure}%
\begin{subfigure}{.49\textwidth}
  \centering
  \includegraphics[width=1\linewidth]{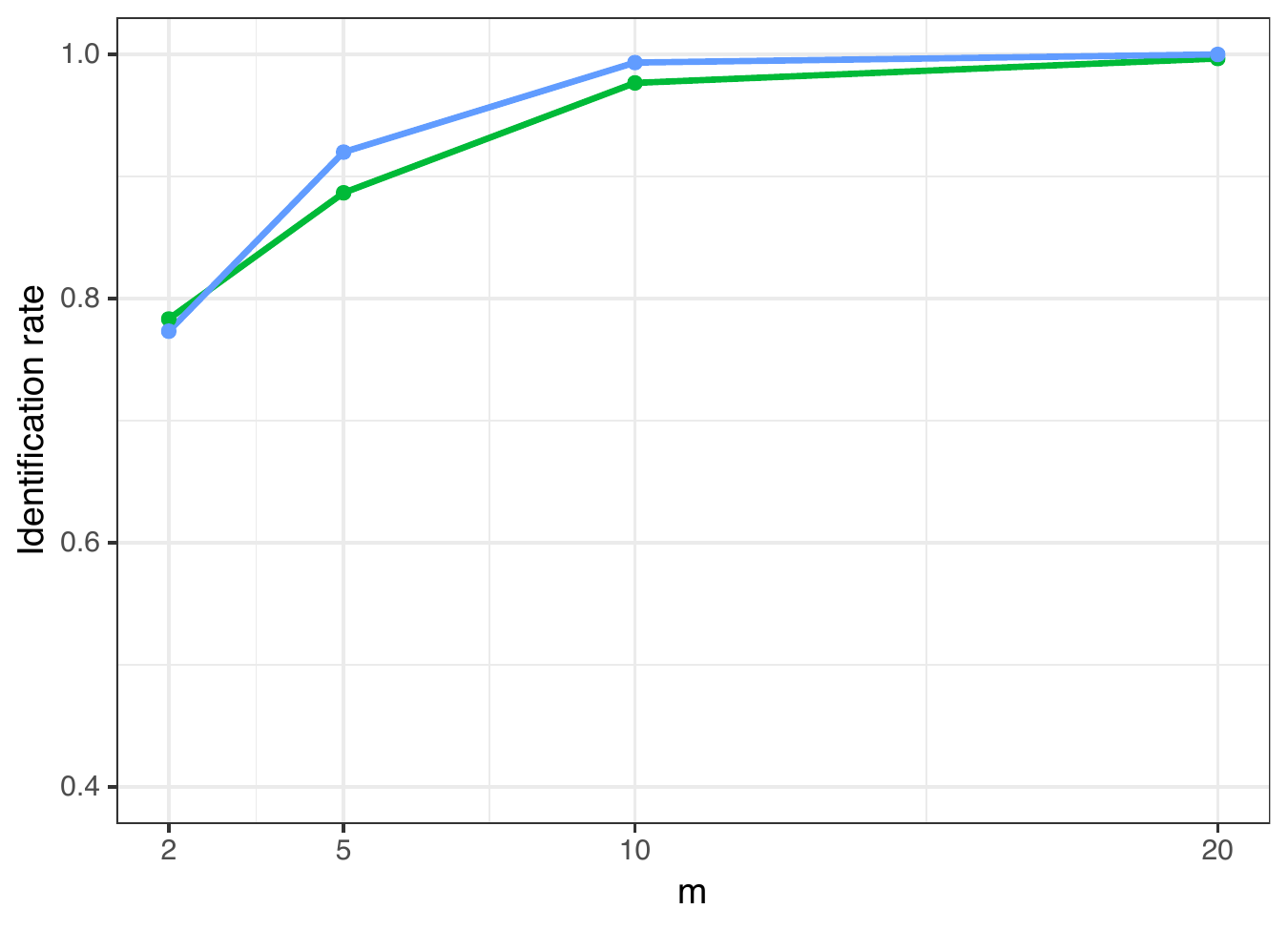}
  \caption{Identification rate}
\end{subfigure}
    \caption{Performance of the oracle estimators $\hrS^*$, $\hPsiX^*$, and $\hPsiY^*$ as a function of $m$ over $100$ replications. (a) Accuracy, the proportion of replications in which the estimate matches the truth exactly. (b) Identification rate for $\hPsiX^*$ and $\hPsiY^*$, the proportion of true indices correctly identified.}
    \label{fig:consistency simul}
\end{figure}

With the same simulation setting, we empirically verify Theorems~\ref{thm:rank consistency} and \ref{thm:index consistency}. Because the targets $r_S$, $\Psi_X$, and $\Psi_Y$ are discrete, we construct the oracle estimators $\hrS^*$, $\hPsiX^*$, and $\hPsiY^*$ defined in the theorems and report two performance metrics over $100$ replications: \emph{accuracy} (the proportion of replications where the estimate exactly matches the true parameter) and, for the index estimators, \emph{identification rate} (the proportion of true indices correctly identified, averaged over replications). Figure~\ref{fig:consistency simul}(a) shows accuracy as a function of $m$. The accuracy of $\hrS^*$ reaches $0.97$ at $m = 5$ and $1.00$ for $m \geq 10$. The accuracy of $\hPsiX^*$ and $\hPsiY^*$ is low at small $m$, since accuracy as defined credits an estimate only when it matches the true set exactly (an estimate of $\{1, 3, 5\}$ for $\Psi_X = \{1, 3, 6\}$ counts as incorrect despite identifying two indices). As $m$ grows, accuracy improves rapidly, approaching $0.99$ and $1.00$ respectively at $m = 20$, consistent with Theorem~\ref{thm:index consistency}. Figure~\ref{fig:consistency simul}(b) shows the identification rate, a less stringent metric: the estimate $\{1, 3, 5\}$ in the example above yields $2/3$. By this measure, $\hPsiX^*$ and $\hPsiY^*$ perform well even at small $m$.

\subsection{Choice of the cutoff margin $\epsilon$}\label{sec:epsilon}

Algorithm~\ref{alg:shared rank} utilizes the asymptotic cutoff $\phi_{c_X}(\lambda_{r_S, S}) \phi_{c_Y}(\gamma_{r_S, S})$. In finite samples, the smallest shared singular value $\sigma_{r_S}(\hQ_S^\top \hU_S)$ fluctuates around its limit and can fall below the cutoff, causing $\hrS$ to be underestimated. A standard remedy is to relax the cutoff to
\[
\phi_{c_X}(\tlamb_{r_S, S}) \phi_{c_Y}(\tgamm_{r_S, S}) - \epsilon
\]
for a nonnegative margin $\epsilon$. A positive $\epsilon$ guards against underestimation at the cost of admitting spurious shared directions. We compare three choices for $\epsilon$ in this appendix and report a simulation study on their relative performance.

\textbf{Choice 1: $\epsilon = 0$.} The cutoff is set to its asymptotic value, ignoring finite-sample fluctuations. This is conservative in that it minimizes the risk of overestimating $r_S$, which is the more damaging error since spurious shared directions corrupt the estimate of $\Sigma_X$.

\textbf{Choice 2: $\epsilon = 1/\sqrt{n_X + n_Y}$.} A heuristic motivated by the $1/\sqrt{n}$ convergence rate of supercritical spiked eigenvectors in the high-dimensional regime \citep[Theorem~5]{paul2007asymptotics}. Convenient but lacking formal justification.

\textbf{Choice 3: parametric bootstrap.} We set $\epsilon$ to the asymptotic standard deviation of the cutoff $\phi_{c_X}(\tlamb_{r_S, S}) \phi_{c_Y}(\tgamm_{r_S, S})$. The randomness in this quantity comes only from the two independent sample eigenvalues $\hlamb_{r_S, S}$ and $\hgamm_{r_S, S}$, since the debiasing function $d$ and the cosine function $\phi_c$ are deterministic. By Theorem~3 of \citet{paul2007asymptotics}, these eigenvalues are asymptotically Gaussian,
\begin{equation}\label{eq:sample eigenvalue distribution}
\hlamb_{r_S, S} \sim N\!\Big(\mu_{c_X}(\lambda_{r_S, S}), \tfrac{1}{n_X} \sigma_{c_X}^2(\lambda_{r_S, S})\Big), \qquad \hgamm_{r_S, S} \sim N\!\Big(\mu_{c_Y}(\gamma_{r_S, S}), \tfrac{1}{n_Y} \sigma_{c_Y}^2(\gamma_{r_S, S})\Big),
\end{equation}
where $\mu_c(\ell) = \ell + c\ell/(\ell - 1)$ and $\sigma_c^2(\ell) = 2\ell^2 (1 - c/(\ell - 1)^2)$. The cutoff's standard deviation has no simple closed form because $d$ and $\phi_c$ are nonlinear, so we approximate it by drawing $1000$ pairs from \eqref{eq:sample eigenvalue distribution} and taking the empirical standard deviation of the resulting cutoff values. The procedure requires only univariate Gaussian sampling, runs in well under a second, and is independent of $n_X$, $n_Y$, and $p$. The true spikes in \eqref{eq:sample eigenvalue distribution} are unknown and replaced by their debiased estimates $\tlamb_{r_S, S}$ and $\tgamm_{r_S, S}$; since estimates for $\Psi_X$ and $\Psi_Y$ are updated at each iteration of Algorithm~\ref{alg:shared rank}, the bootstrap standard deviation is recomputed at every step.

\begin{table}[!h]
\centering
\begin{tabular}{cl ccccc l}
\hline
Case & Dataset & $p$ & $n$ & $r_S$ & $d$ & $\sigma(Q_D^\top U_D)$ & Eigenvalues \\
\hline
\multirow{2}{*}{1} & $X$ & $500$ & $100$ & $7$ & $5$ & $\{0.7, \ldots, 0.1\}$ & $\log \lambda_i$ from $3$ to $2$ \\
                   & $Y$ & $500$ & $300$ & $7$ & $8$ & $\{0.7, \ldots, 0.1\}$ & $\log \gamma_j$ from $3.5$ to $1.5$ \\
\hline
\multirow{2}{*}{2} & $X$ & $300$ & $100$ & $10$ & $2$ & $\{0.5, \ldots, 0.1\}$ & $\log \lambda_i$ from $3.5$ to $1.3$ \\
                   & $Y$ & $300$ & $300$ & $10$ & $2$ & $\{0.5, \ldots, 0.1\}$ & $\log \gamma_j$ from $4$ to $1.5$ \\
\hline
\multirow{2}{*}{3} & $X$ & $1000$ & $500$ & $5$ & $12$ & $\{0.8, \ldots, 0.1\}$ & $\log \lambda_i$ from $3.5$ to $1.5$ \\
                   & $Y$ & $1000$ & $1000$ & $5$ & $15$ & $\{0.8, \ldots, 0.1\}$ & $\log \gamma_j$ from $3.5$ to $1.5$ \\
\hline
\end{tabular}
\caption{Simulation parameters for comparing the three $\epsilon$ selection strategies. The singular values of $Q_D^\top U_D$ and the log-eigenvalues are equidistant sequences between the listed endpoints.}
\label{tab:epsilon simul}
\end{table}

We compare the three choices through simulations in three cases summarized in Table~\ref{tab:epsilon simul}. In the table, $n$ and $d$ denote the sample size and distinct dimension respectively. The singular values of $Q_D^\top U_D$ and the logarithms of eigenvalues are taken as equidistant sequences. We choose relatively large singular values for $Q_D^\top U_D$ to provide a clear contrast between the three $\epsilon$ strategies, even though such values are less likely in genuinely high-dimensional settings. For each case, $\Psi_X$ and $\Psi_Y$ are chosen so that the shared eigenvectors are positioned across the spiked eigenvalue spectrum; for instance, $\Psi_X = \{1, 3, 4, 5, 6, 8, 10\}$ and $\Psi_Y = \{1, 2, 4, 6, 7, 10, 12\}$ in Case~1.

\begin{figure}[!h]
\begin{subfigure}{.33\textwidth}
  \centering
  \includegraphics[width=1\linewidth]{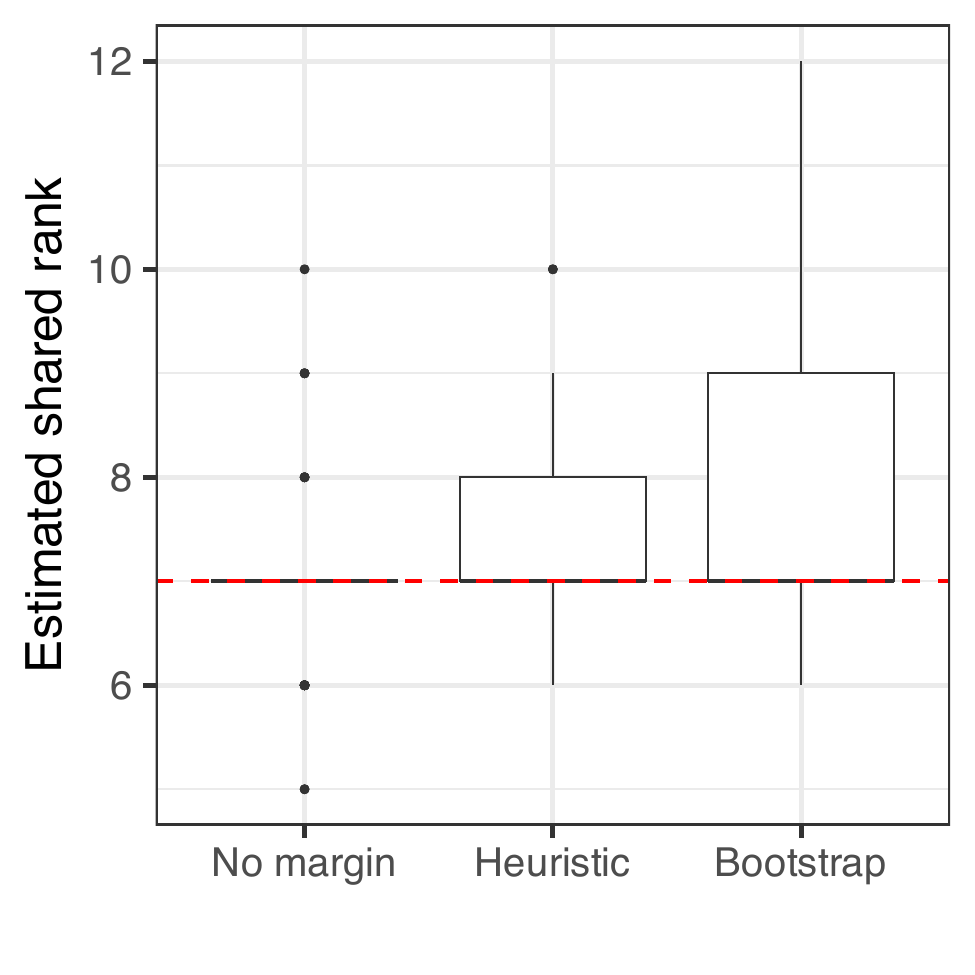}
  \caption{Case 1}
\end{subfigure}%
\begin{subfigure}{.33\textwidth}
  \centering
  \includegraphics[width=1\linewidth]{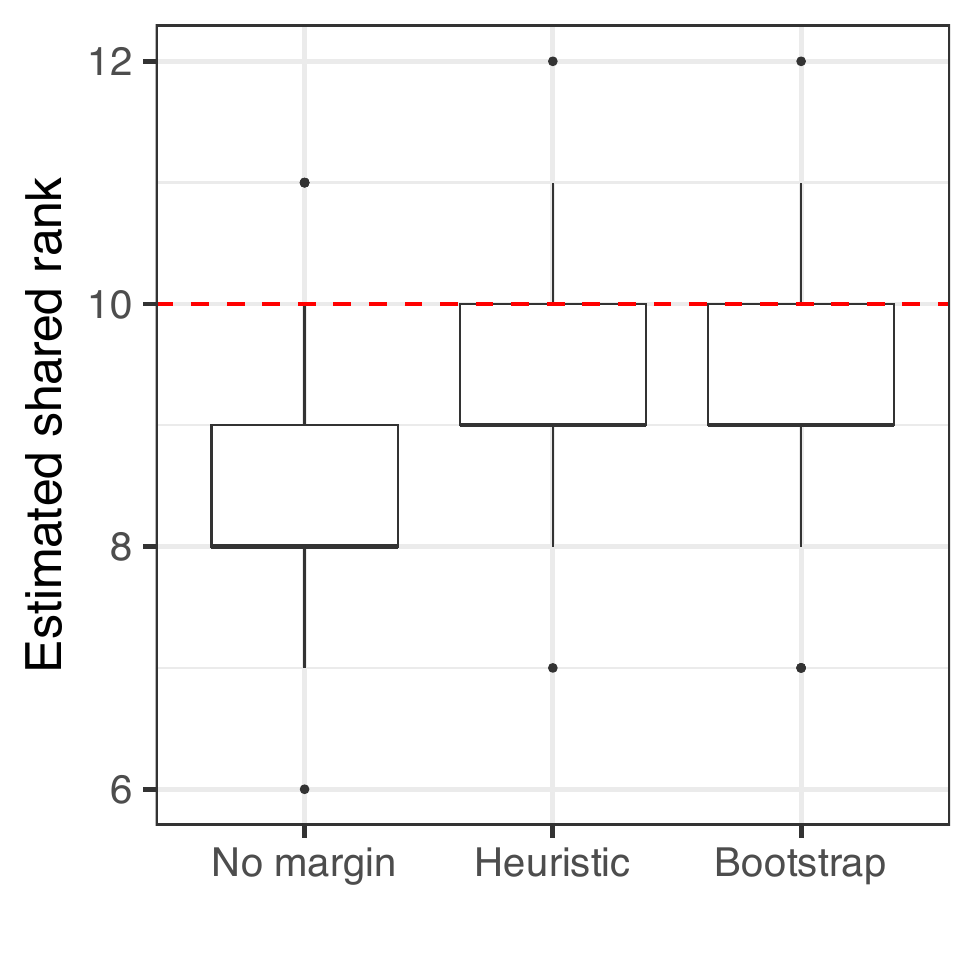}
  \caption{Case 2}
\end{subfigure}
\begin{subfigure}{.33\textwidth}
  \centering
  \includegraphics[width=1\linewidth]{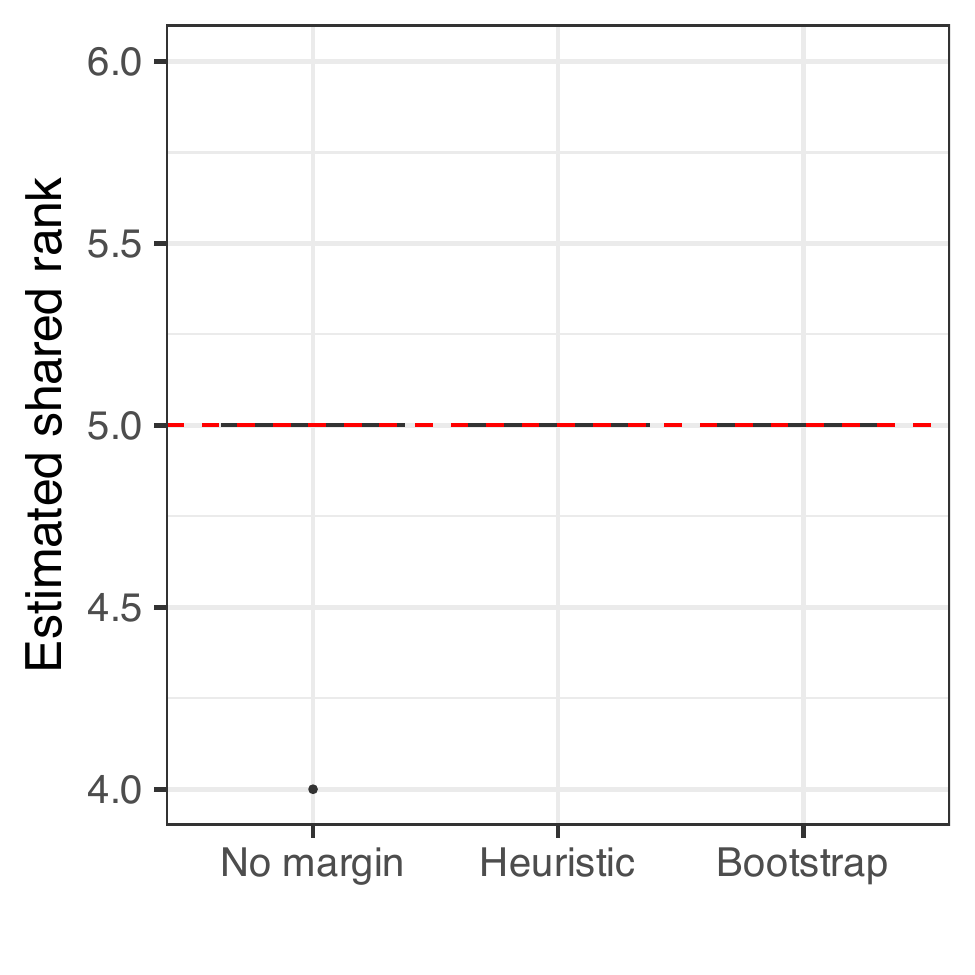}
  \caption{Case 3}
\end{subfigure}
\caption{Boxplots of the estimated shared subspace rank over $100$ replications for the three $\epsilon$ selection strategies ($\epsilon = 0$, $\epsilon = 1/\sqrt{n_X + n_Y}$, and parametric bootstrap), across the three simulation cases in Table~\ref{tab:epsilon simul}. The dashed red line in each panel marks the true shared rank.}
\label{fig:eps simul}
\end{figure}

Figure~\ref{fig:eps simul} shows the estimated shared subspace rank over $100$ replications for each of the three $\epsilon$ selection strategies. The dashed red line in each panel marks the true shared rank for the corresponding case. No single strategy dominates across all cases. In Case~1, where both datasets are strongly high-dimensional ($c_X = 5$, $c_Y = 5/3$), $\epsilon = 0$ gives the most accurate estimates and positive margins lead to overestimation. In Case~2, the shared rank is large relative to the distinct ranks, and Algorithm~\ref{alg:shared rank} underestimates the rank at $\epsilon = 0$ despite the smaller dimension; the heuristic and bootstrap margins both correct the underestimation. In Case~3, where dimension and sample sizes are both large, all three strategies recover the true rank accurately, even with $\sigma_1(Q_D^\top U_D) = 0.8$, an unusually large value for genuinely high-dimensional settings.

\subsection{Instability of selective pooling}\label{sec:appendix-total vs selective}

We provide empirical evidence for the finite-sample instability of the selective estimators $\hP_{X,S} = \hQ_S\hQ_S^\top$ and $\hP_{Y,S}=\hU_S\hU_S^\top$, motivating the use of the total pooling estimator $\bar{P}_{r_S}$.

We set $p = 500$, $n_X = n_Y = 600$, $r_S = 3$, $d_X = d_Y = 4$, with eigenvalues $\Lambda = \mathrm{diag}(19, 16, 13, 9, 7, 5, 3)$ and $\Gamma = \mathrm{diag}(18, 15, 12, 10, 8, 6, 3)$, index sets $\Psi_X = \{1, 3, 5\}$ and $\Psi_Y = \{1, 3, 4\}$, and noise levels $\tau^2 = \eta^2 = 1$.

\begin{figure}[!h]
\centering
\begin{subfigure}{.5\textwidth}
\centering
\includegraphics[width=\linewidth]{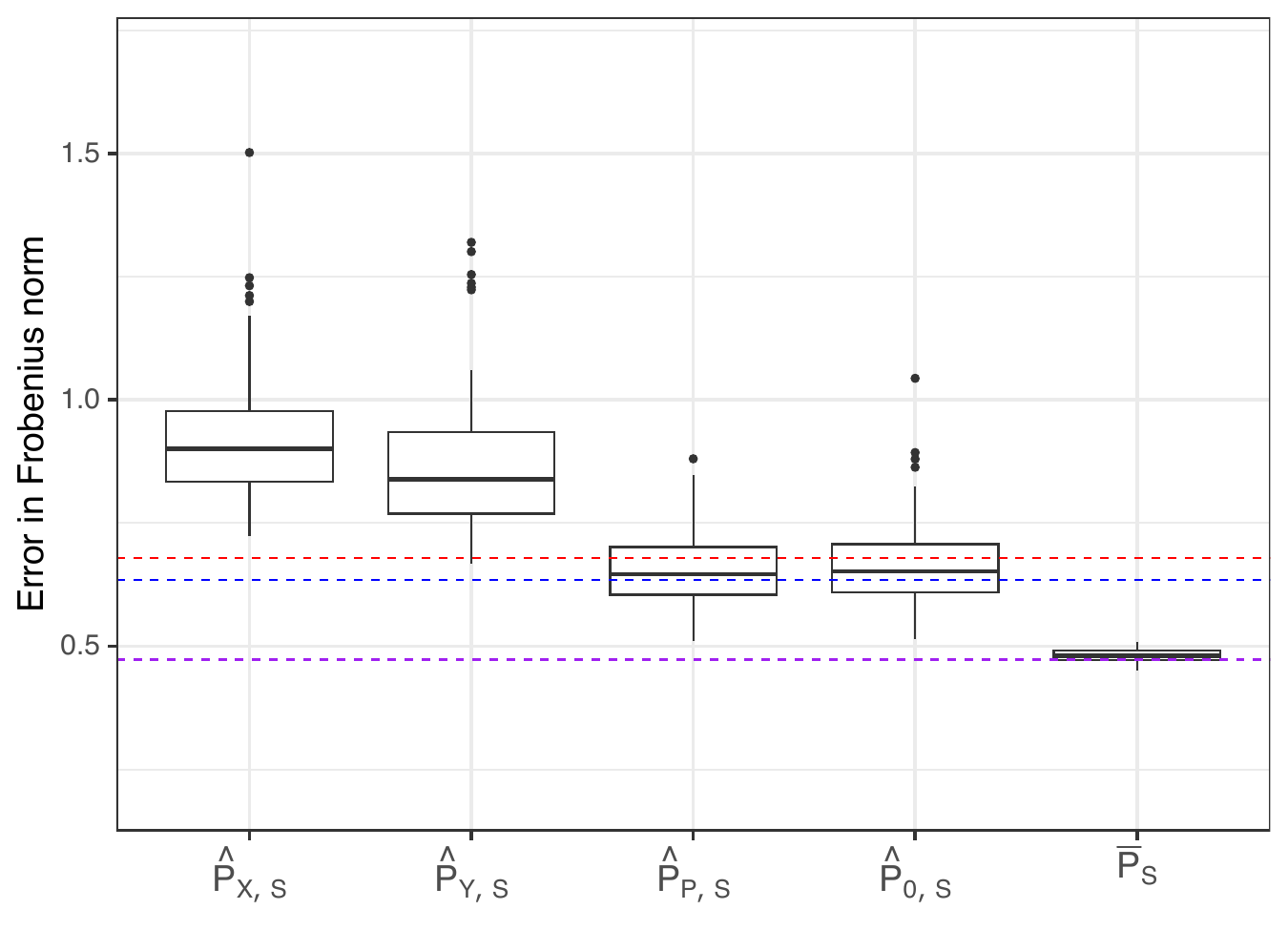}
\caption{Full model}
\end{subfigure}%
\begin{subfigure}{.5\textwidth}
\centering
\includegraphics[width=\linewidth]{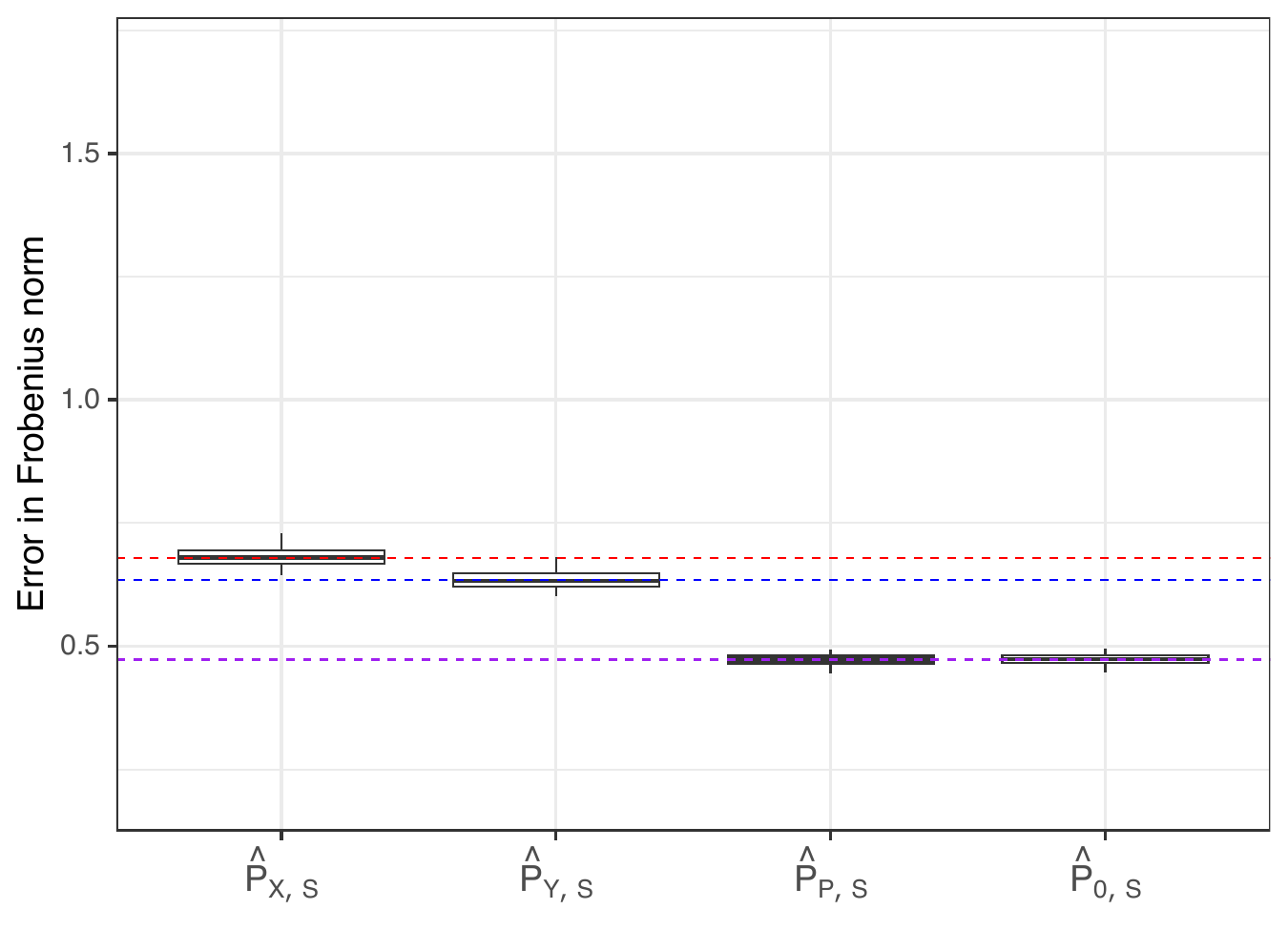}
\caption{Reduced model}
\end{subfigure}
\caption{Frobenius errors of the selective and total pooling estimators of $P_S$, over $100$ replications. (a) Full PSS model. (b) Reduced model with no distinct components. Horizontal dotted lines (red, blue, purple) mark the asymptotic Frobenius errors of $\hP_{X,S}$, $\hP_{Y,S}$, and $\hP_{P,S}$ respectively.}
\label{fig:total vs selective}
\end{figure}

Figure~\ref{fig:total vs selective}(a) shows the Frobenius error in estimating $P_S$ over $100$ replications, for five estimators: $\hP_{X,S}$, $\hP_{Y,S}$, the simple pooling $\hP_{P,S} = \tfrac{1}{2}(\hP_{X,S} + \hP_{Y,S})$, the projection pooling $\hP_{0,S} = \Pi_{r_S}(\hP_{X,S} + \hP_{Y,S})$, and the total pooling $\bar{P}_{r_S}$. Horizontal dotted lines in red, blue, and purple mark the asymptotic Frobenius errors of $\hP_{X,S}$, $\hP_{Y,S}$, and $\hP_{P,S}$, respectively. The total pooling estimator $\bar{P}_{r_S}$ achieves its asymptotic error rate and substantially outperforms $\hP_{P,S}$ and $\hP_{0,S}$, consistent with Theorem~\ref{thm:Frobenius performance}. The remaining estimators show empirical errors well above their asymptotic limits, indicating finite-sample instability.

To isolate the source of this instability, we repeat the simulation with the distinct components removed:
\[
\Sigma_{X, \text{reduced}} = Q_S \Lambda_S Q_S^\top + \tau^2 I_p, \qquad \Sigma_{Y, \text{reduced}} = U_S \Gamma_S U_S^\top + \eta^2 I_p,
\]
so that the sample shared eigenvectors $\hQ_S$ and $\hU_S$ are not perturbed by the within-signal interference from $\hQ_D$ and $\hU_D$. Figure~\ref{fig:total vs selective}(b) shows the corresponding Frobenius errors. Under the reduced model, all four selective estimators attain their asymptotic error rates. The contrast with panel (a) identifies within-signal perturbation as the source of the finite-sample instability, and explains the advantage of $\bar{P}_{r_S}$ in the full model.

\subsection{Optimality of the pooling weight}\label{sec:appendix-pooling weight}

Theorem~\ref{thm:Frobenius performance} motivates the choice of pooling weight that minimizes the asymptotic Frobenius error of $\hP_S$. This appendix demonstrates the effectiveness of the optimization through simulated data.

We set $r_S = d_X = d_Y = 4$, $\Psi_X = \{1, 3, 4, 6\}$, $\Psi_Y = \{1, 2, 5, 6\}$, and $\tau^2 = \eta^2 = 1$. The dimension, sample sizes, and eigenvalues are varied across two cases, summarized in Table~\ref{tab:pooling weight simul}: Case~1 has $n_X \ll n_Y$ but larger eigenvalues in the target dataset, while Case~2 has $n_X = n_Y$ but larger eigenvalues in the background dataset. These two regimes correspond to the two distinct ways the optimal pooling weight can deviate from a sample-size-based weight.

\begin{table}[!h]
\centering
\begin{tabular}{cl ccc}
\hline
Case & Dataset & $p$ & $n$ & Eigenvalues \\
\hline
\multirow{2}{*}{1} & $X$ & $500$ & $300$ & $\log \lambda_i$ from $4.5$ to $2.5$ \\
                   & $Y$ & $500$ & $1000$ & $\log \gamma_j$ from $3.5$ to $1.5$ \\
\hline
\multirow{2}{*}{2} & $X$ & $500$ & $300$ & $\log \lambda_i$ from $3.5$ to $1.5$ \\
                   & $Y$ & $500$ & $300$ & $\log \gamma_j$ from $4.7$ to $2.5$ \\
\hline
\end{tabular}
\caption{Simulation parameters for the pooling weight optimization study. The log-eigenvalues are equidistant sequences between the listed endpoints.}
\label{tab:pooling weight simul}
\end{table}

\begin{figure}[!h]
\centering
\begin{subfigure}{.45\textwidth}
\centering
\includegraphics[width=\linewidth]{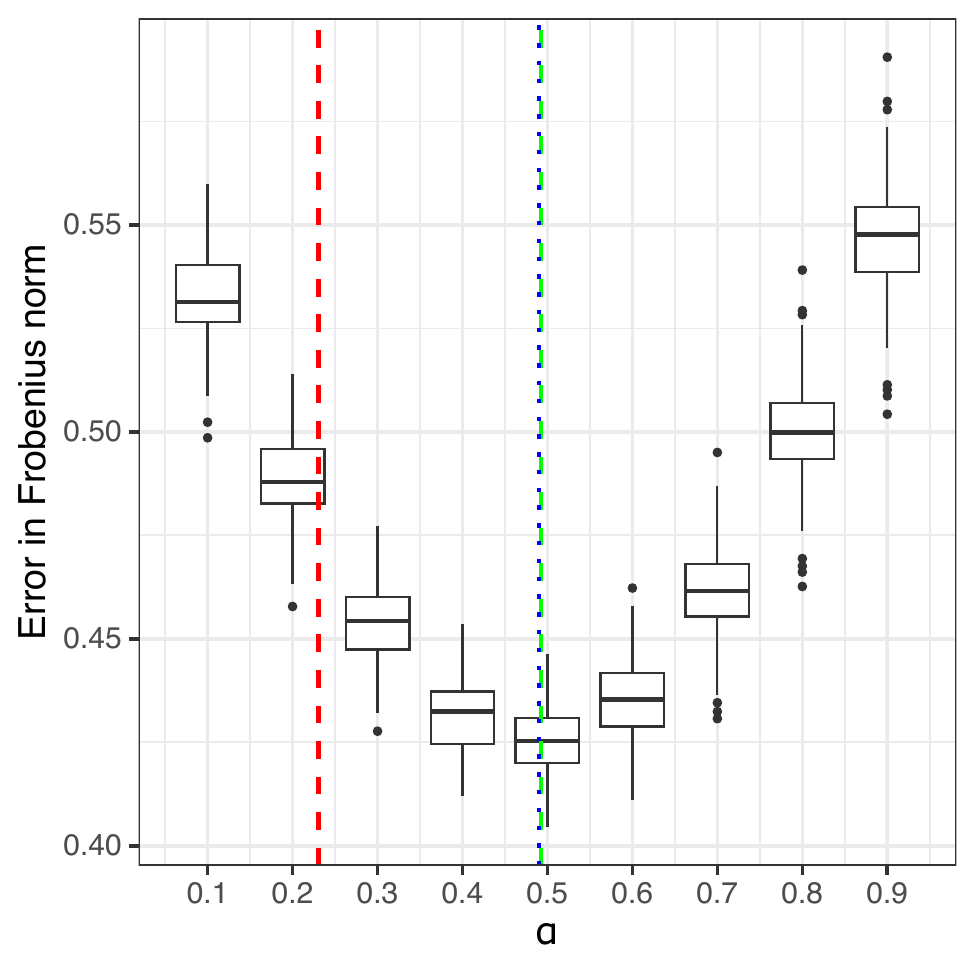}
\caption{$P_S$ (Case 1)}
\end{subfigure}%
\begin{subfigure}{.45\textwidth}
\centering
\includegraphics[width=\linewidth]{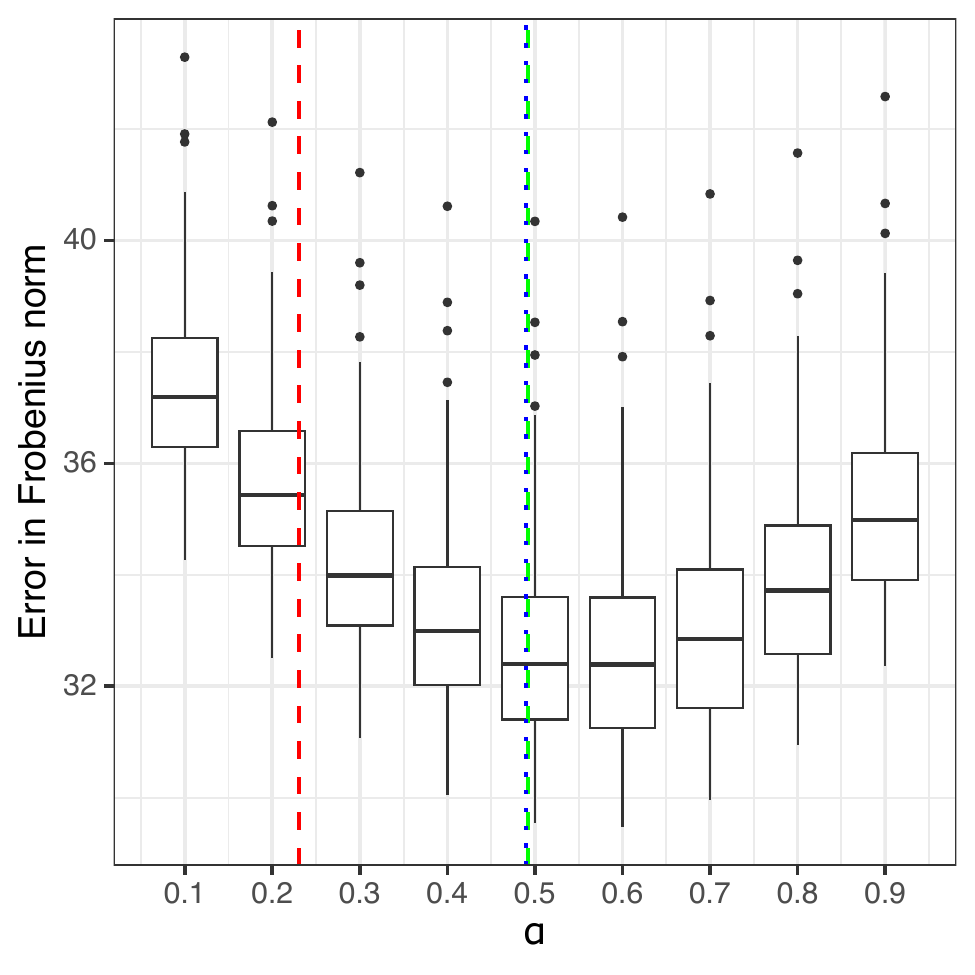}
\caption{$\Sigma_X$ (Case 1)}
\end{subfigure}

\begin{subfigure}{.45\textwidth}
\centering
\includegraphics[width=\linewidth]{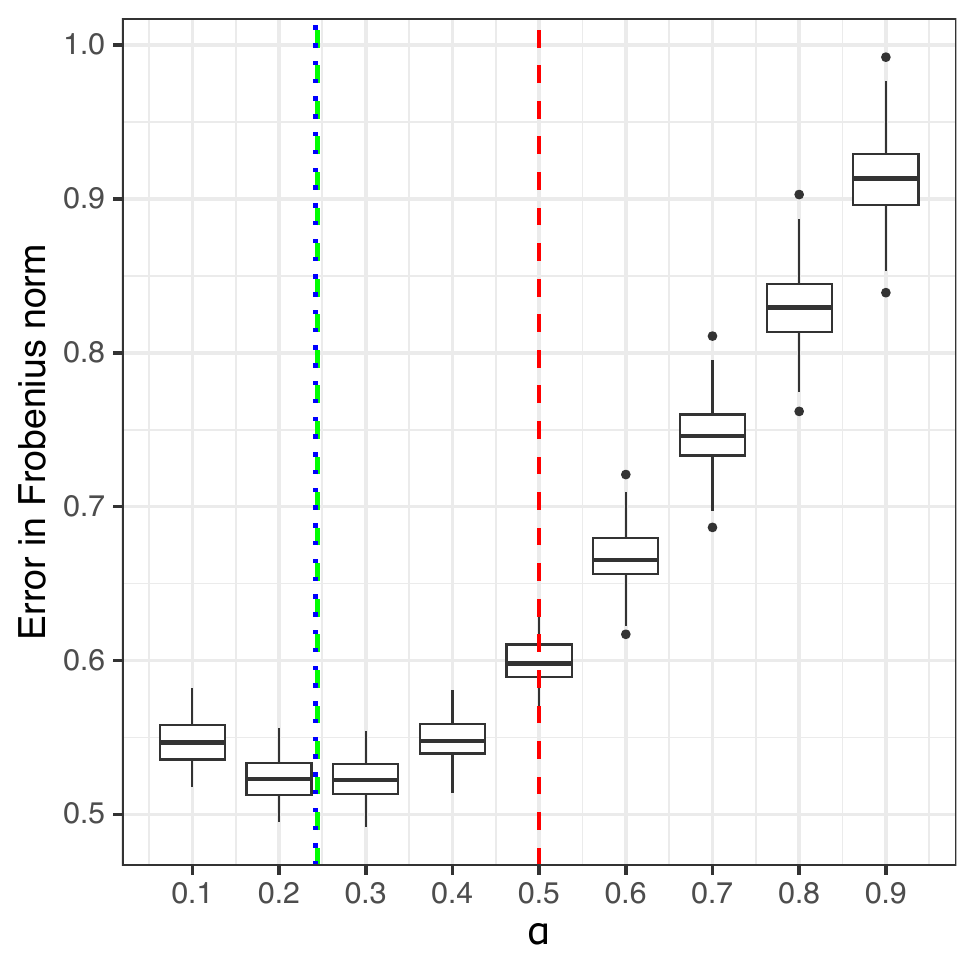}
\caption{$P_S$ (Case 2)}
\end{subfigure}%
\begin{subfigure}{.45\textwidth}
\centering
\includegraphics[width=\linewidth]{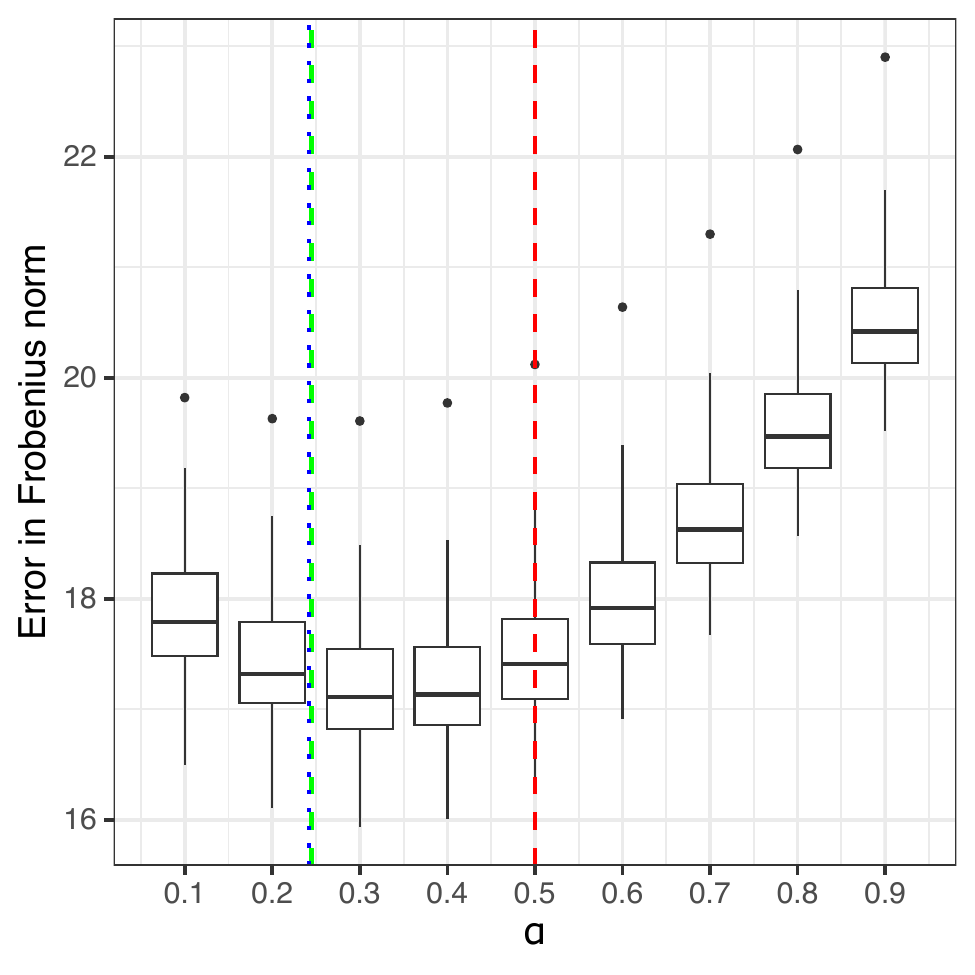}
\caption{$\Sigma_X$ (Case 2)}
\end{subfigure}
\caption{Boxplots of Frobenius errors in estimating $P_S$ and $\Sigma_X$ over $100$ replications, as a function of the pooling weight $\alpha$. The red, green, and blue dashed lines mark $\tilde\alpha = n_X/(n_X + n_Y)$ (sample-size-based), $\alpha^*$ (population optimum), and $\hat\alpha$ (plug-in estimate), respectively.}
\label{fig:pooling weight case}
\end{figure}

Figure~\ref{fig:pooling weight case} shows boxplots of the Frobenius error in estimating $P_S$ and $\Sigma_X$ over $100$ replications, for $\alpha$ values from $0.1$ to $0.9$. The red dashed line marks the sample-size-based weight $\tilde\alpha = n_X / (n_X + n_Y)$; the green and blue dashed lines mark the population optimum $\alpha^*$ and its plug-in estimate $\hat\alpha$. Two observations stand out. First, $\hat\alpha$ closely tracks $\alpha^*$ in both cases, with the green and blue lines nearly overlapping. Second, the Frobenius error is minimized near $\alpha^*$ for both $P_S$ and $\Sigma_X$, while the sample-size-based weight $\tilde\alpha$ yields suboptimal estimates in both cases. The benefit of optimization is most apparent in Case~2, where $\tilde\alpha = 1/2$ ignores the eigenvalue asymmetry between the two datasets.

\newpage

\subsection{Shared rank estimation in degenerate PSS model}\label{sec:appendix-degenerate PSS}

 This experiments considers the two degenerate endpoints of the PSS model: the \emph{non-shared} case ($r_S = 0$, so each covariance matrix consists only of distinct spikes), and the \emph{fully-shared} case ($d_X = d_Y = 0$, so each covariance matrix consists only of shared spikes, with true rank $r_S = 5$). In the non-shared case, we additionally require $Q_D^\top U_D = 0_{d_X\times d_Y}$, ruling out spurious overlap between the distinct subspaces. The distinct components in the non-shared setting are taken from the mixed configuration~(b) of Section~\ref{subsec:est-sim}. Each setting is replicated $100$ times.

Table~\ref{table:no_fully-shared-rank} reports the average estimated shared rank. In the non-shared case, PSS and CDE both return zero across all sample sizes, correctly recognizing the absence of a shared subspace. MgCov returns a strictly positive rank in every case, indicating that it cannot detect the absence of shared structure. PerPCA selects $r_S = 15$, the upper end of its grid, in every replication. In the fully-shared case, the situation is reversed: PSS recovers the true rank as the sample size grows; MgCov is exact (this regime matches its model assumptions); CDE largely fails to detect the shared structure; and PerPCA again fails to recover the true rank.

The contrast between the two cases is the substantive finding. PSS is the only method that handles both endpoints correctly, returning zero when no shared structure exists and the true rank when all structure is shared. MgCov succeeds in the fully shared case but cannot recognize the non-shared case—a meaningful limitation, since whether a shared structure exists is itself something an analyst needs to determine before applying any shared-subspace method. PSS, therefore, serves as a diagnostic tool for the existence of a shared subspace, as well as for estimating it when present.

\begin{table}[htbp]
    \centering
    \resizebox{0.8\linewidth}{!}{%
      \begin{tabular}{llcccc}
      \hline
      \multirow{2}{*}{Setting} & \multirow{2}{*}{$(p, n_X, n_Y)$} & \multicolumn{4}{c}{ Estimated shared rank} \\
      \cline{3-6}
       & & PSS & CDE &  MgCov & PerPCA \\
      \hline
      \multirow{3}{*}{Non-shared} & $(200, 50, 200)$    & 0.00 (0.00) & 0.00 (0.00) & 7.51 (0.06) & 15.00 (0.00) \\
                                  & $(500, 125, 500)$   & 0.00 (0.00) & 0.00 (0.00) & 7.85 (0.04) & 15.00 (0.00) \\
                                  & $(1000, 250, 1000)$ & 0.01 (0.01) & 0.00 (0.00) & 7.97 (0.02) & 15.00 (0.00) \\
      \hline
      \multirow{3}{*}{Fully-shared}  & $(200, 50, 200)$   & 3.78 (0.05) & 0.07 (0.03) & 5.00 (0.00) & 15.00 (0.00) \\
                                    & $(500, 125, 500)$   & 4.62 (0.04) & 0.00 (0.00) & 5.00 (0.00) & 15.00 (0.00) \\
                                    & $(1000, 250, 1000)$ & 4.99 (0.01) & 0.52 (0.05) & 5.00 (0.00) & 15.00 (0.00) \\
      \hline
      \end{tabular}
    }
    \caption{Shared rank estimation in two degenerate cases of the PSS model: the non-shared case (true $r_S = 0$) and the fully shared case (true $r_S = 5$).}
    \label{table:no_fully-shared-rank}
\end{table}

\bibliographystyle{unsrtnat}

\bibliography{references}

\makeatletter\@input{xx_article.tex}\makeatother